\documentclass[aps, prd, amsmath, floats, floatfix, twocolumn, nofootinbib, superscriptaddress, showpacs]{revtex4-1}
\usepackage{graphicx}
\usepackage{multirow}
\usepackage{color}
\usepackage{latexsym}
\usepackage{amsfonts}
\usepackage{comment}
\usepackage{bm}
\usepackage[colorlinks]{hyperref}
\graphicspath{{figures/}} 

\newcommand{\be}{\begin{eqnarray}}
\newcommand{\ee}{\end{eqnarray}}

\begin{document}

	\title{Gravitational waves and electromagnetic radiation from charged black hole binaries}

	\author{Carlos~A.~Benavides-Gallego}
	\email[]{cabenavidesg20@shao.ac.cn}
	\affiliation{Shanghai Astronomical Observatory, 80 Nandan Road, Shanghai 200030, P. R. China}

	\author{Wen-Biao Han}
	\email[Corresponding author: ]{wbhan@shao.ac.cn}
	\affiliation{Shanghai Astronomical Observatory, 80 Nandan Road, Shanghai 200030, P. R. China}
	\affiliation{School of Fundamental Physics and Mathematical Sciences, Hangzhou Institute for Advanced Study, UCAS, Hangzhou 310024, China}
	\affiliation{School of Astronomy and Space Science, University of Chinese Academy of Sciences, Beijing 100049, P. R. China}
	\affiliation{International Centre for Theoretical Physics Asia-Pacific, Beijing/Hangzhou 310024, China}
    \affiliation{Shanghai Frontiers Science Center for Gravitational Wave Detection, 800 Dongchuan Road, Shanghai 200240, China}
	\date{\today}

    \begin{abstract}
    It is still an open issue if astrophysical black holes have electric charges or not. In this work, we analytically calculate gravitational and electromagnetic waveforms in the frequency domain for charged black hole binaries during the inspiral phase. In addition to the well-known $f^{-7/6}$ waveforms, we also get a $-11/6$ power law gravitational wave component. The phase of waveforms for charged binary is fully derived. In the case of electromagnetic counterparts, we focus on the electromagnetic dipole radiation, but we include the quadrupole contribution to complete our discussion. We also obtain the chirp property of the electromagnetic waves. In the case of dipole radiation, the frequency-domain waves are proportional to $f^{-7/6}$, while $f^{-1/6}$ appears in the quadrupole contribution. The frequency-domain waveforms can be used to estimate the charges of black holes in the current gravitational wave observations.
	\end{abstract}
	
	\maketitle
	

	\section{Introduction}

	The detection of gravitational waves by LIGO and Virgo scientific collaborations~\cite{LIGOScientific:2016aoc, LIGOScientific:2016emj, LIGOScientific:2018mvr} has opened the possibility of exploring and understanding the nature of gravity in the strong-field regime, giving us the chance to test general relativity (GR) and compare its predictions with alternative theories and observations~\cite{Berti:2015itd}. Nowadays, the observation of gravitational waves (GWs) is constrained to the frequency range of $10-10^3$Hz. Therefore, ground-based detectors can not measure GW signals if the frequency oscillates between $10^{-4}-10^{-1}$Hz, where the astrophysical signals reside. Nevertheless, space-based observatories, such as LISA~\cite{Danzmann:1997hm}, TianQin~\cite{TianQin:2015yph} and Taiji~\cite{Hu:2017mde} will improve the accuracy and range of observations, opening the window to the low-frequency detection of GWs.

	One of the sources of GW signals detectable using space-based observatories is a binary system formed by a stellar-mass compact object, such as black holes (BHs) or neutron stars (NS), orbiting a supermassive black hole (SMBH). When the mass ratio of these systems oscillates between $10^{-7}$ and $10^{-4}$, we use the term \textit{extreme mass ratio inspirals} (EMRIs) to name them. It is well-known that EMRIs are suitable for investigating the mass, the spin, the electric charge, and the strong-field physics in the vicinity of BHs~\cite{Amaro-Seoane:2007osp, Babak:2017tow, Berry:2019wgg, Fan:2020zhy, Zi:2021pdp}. On the other hand, according to GR, any astrophysical black hole can be described by three external parameters: mass, angular momentum, and electric charge. This follows as a consequence of the well-known \textit{no-hair theorem}~\cite{Israel:1967wq, Israel:1967za, Carter:1971zc}. Hence, from the observational point of view, one expects that the reason behind the multi-messenger\footnote{This term refers to observations of electromagnetic radiation, gravitational waves, neutrinos, and cosmic rays~\cite{Zajacek:2019kla}.} experiments is to determine these external parameters. Nevertheless, only the black hole mass and its angular momentum have been taken into account, while the electric charge, on the other hand, is usually neglected and set equal to zero. As claimed by M. Zaja\v{c}ek and A.~Tursunov, ``\textit{this assumption is supported by arguing that the presence of plasma around astrophysical black holes leads to prompt discharging}''~\cite{Zajacek:2019kla}. 
    The presence of charge in compact objects is still in debate, and the question \textit{how could black holes get charged?} has been considered by several authors~\cite{Eddington:1926, Wald:1974np, Gibbons:1975kk, Bally:1978, Zajacek:2018ycb, Zajacek:2018vsj}. In Ref.~\cite{Eddington:1926}, for example, to prevent the separation of electrons and protons in the stellar atmosphere, Eddington suggested that stars should have a small positive charge. In Ref.~\cite{Wald:1974np}, Wald proposed a relativistic mechanism that supports the existence of charged black holes. According to Wald, when one immerses a rotating black hole in a uniform magnetic field, an electric field is induced due to the twisting of magnetic field lines, implying that a non-zero charge is conceivable. The value of the induced electric charge is proportional not only to the strength of the magnetic field but also to the black hole's spin~\cite{Zajacek:2019kla, Zajacek:2018ycb, Zajacek:2018vsj}. The sign of the electric charge induced via the Wald mechanism depends on the orientation of magnetic field lines in relation to the black hole spin. For example, the black hole would have positive charge if the magnetic field is parallel to the rotation axis of the black hole. In this sense, since a certain degree of alignment between the accretion flow angular momentum and the black hole spin is expected, the charge of astrophysical black holes tends to be positive~\cite{Zajacek:2019kla}. Later, in 1978, John Bally and Harrison showed that any macroscopic body in the universe, such as stars, galaxies, and black holes, are positively charged with the charge-to-mass ratio of approximately $100$ Coulombs per Solar mass~\cite{Bally:1978}.  
    
    Recently, there has been an increasing interest in charged black holes, see Refs.~\cite{Zajacek:2018vsj,Kim:2000gy,Lee:2000tm,Bozzola:2019aaw, Liu:2020cds,Liu:2020vsy,Liu:2020bag,Christiansen:2020pnv,Bozzola:2020mjx, Bozzola:2021elc,Wang:2021vmi,Liu:2022cuj,Luna:2022udb,Karas:2017wre, Kopacek:2018rkb, Levin:2018mzg} and references therein. In Ref.~\cite{Zajacek:2018vsj}, M. Zaja\v{c}ek et al. used observations of the Galactic center black hole Sgr A* to constrain its charge. They also used their results to analyze two of the most interesting astrophysical consequences of slightly charged black holes: the effect on the gamma-ray bursts\footnote{Also called X-ray \textit{bremsstrahlung} by M. Zaja\v{c}ek et al. in Ref.~\cite{Zajacek:2018vsj}.} (GRBs) profile and the effect on the position of the innermost stable circular orbit (ISCO). Although a small charge does not affect the space-time structure drastically, the authors were able to show that it may be of relevance for the plasma dynamics close to the Galactic center black hole/supermassive black holes. Moreover, the authors concluded that the charge and the associated electromagnetic (EM) signal could be crucial for plunges of neutron stars into supermassive black holes or black hole-neutron star mergers~\cite{Wang:2021vmi, Karas:2017wre, Kopacek:2018rkb, Levin:2018mzg}.
    
    In Refs.~\cite{Kim:2000gy, Lee:2000tm}, the authors investigated the charge and magnetic flux on rotating black holes, showing that black holes and magnetars carry similar charges in sign and magnitude. Moreover, in the collapsar/hypernova scenario of gamma-ray bursts, the results indicate that the central electric charge and the associated magnetic flux remain continuous. On the other hand, regarding the extraction of rotational energy, the authors found that this process will continue, provided the magnetic field remains supported by the surrounding magnetized matter. 
    
    In Refs.~\cite{Bozzola:2019aaw, Liu:2020cds, Liu:2020vsy, Liu:2020bag, Christiansen:2020pnv, Bozzola:2020mjx, Bozzola:2021elc,Wang:2021vmi,Liu:2022cuj,Luna:2022udb}, the authors consider charged black holes to investigate gravitational-wave physics. In Ref.~\cite{Bozzola:2019aaw}, for example, G.~Bozzola and V.~Paschalidis developed an initial data formalism valid for general relativistic simulations of binary systems with electric charge and linear and angular momenta. As claimed by the authors, the formalism is useful for simulating the dynamical evolution of the ultrarelativistic head-on collision, the quasicircular or eccentric inspiral, and the merger of two black holes~\cite{Bozzola:2020mjx, Bozzola:2021elc}. 
    
    L.~Liu et al. studied the case of BH binaries with electric and magnetic charges in circular and elliptical orbits on a cone in Refs.~\cite{Liu:2020cds, Liu:2020vsy, Liu:2020bag}. First, the authors considered a BH binary system formed by non-rotating dyonic black holes. Then, using the Newtonian approximation with radiation reactions, they calculated the total emission rate of energy and angular momentum generated by the gravitational and EM radiation. In the case of circular orbits, they showed that electric and magnetic charges significantly suppress the merger times of binaries. On the other hand, when considering elliptical orbits, they showed that the emission rates of energy and angular momentum produced by the gravitational and EM radiation have the same dependence on the conic angle for different orbits.
    
    Finally, in Ref.~\cite{Christiansen:2020pnv}, Christiansen et al. investigated the emission of GWs by systems involving charged BHs whose charge corresponds to some dark-charge. The authors explain that this kind of BHs can be created in the early universe by self-interacting dark matter (DM) models. The main idea of their work is to ``\textit{investigate some observational consequences of compact objects beyond those well-captured by the employed templates}''~\cite{Christiansen:2020pnv}. To do so, they begin by considering Keplerian orbits where the emission comes mainly from the EM dark-charge dipole contribution, which they use later to obtain the time evolution of the orbital parameters in the Newtonian approximation. In that work, the authors show that a good approximation for both EM and GW-dominated emissions (in the LIGO/Virgo sensitivity range) can be obtained by considering circular orbits at the time of the merger.

    In the manuscript, we investigate the EM radiation of a binary system formed by charged black holes. In a previous paper~\cite{Benavides-Gallego:2021the}, we studied the EM radiation of a binary system immersed in a uniform magnetic field using a toy model proposed by C.~Palenzuela et al. in Ref.~\cite{Palenzuela:2009hx}. Following a similar philosophy, we obtain the EM waveform radiated by the system during the inspiral phase. Nevertheless, we use the quasicircular approximation derived in Ref.~\cite{Christiansen:2020pnv} by Christiansen et al. We organize the paper as follows. In Sec.~\ref{SecII}, we follow Ref.~\cite{Goldstein:1980} to discuss and obtain the Keplerian orbits for a system of two point-masses with electric charges. In Secs.~\ref{SecIII} and \ref{SecIV}, we review the gravitational and EM radiation, the angular momentum emission, and the evolution of the orbital parameters, following previous results in the literature. Then, in Sec.~\ref{SecV}, we compute the GW and EM waves. In Sec.~\ref{SecVII}, we obtain the Fourier transform of the EM wave. Finally, in Sec.\ref{SecVIII}, we discuss our results.  
    In the manuscript, we denote vectors using bold letters and scalar with normal letters. The time-average of a quantity $A$ is denoted by $\overline{A}$. On the other hand, we choose CGS units, where the electric constant $k_e=1$~\cite{Landau:1975pou}. Following Ref.~\cite{Maggiore:2007ulw}, we keep $G$, $c$ and $k_e$ in the expressions, with the exception of Sec.~\ref{SecV} and figures, where we use dimensionless units, see Appendix~\ref{A0}.

	\section{Keplerian motion \label{SecII}}
	
	The problem of two bodies moving under the influence of a central force can be analyzed using the Lagrangian formulation~\cite{Goldstein:1980}. Therefore, we devote this section to the Keplerian orbits of two point-particles with masses $m_1$ and $m_2$ and charges $Q_1$ and $Q_2$, respectively. To obtain the equations of motion, we consider a central force given by a function $\mathcal{U}$ containing the gravitational and electric potentials. It is important to remark that $\mathcal{U}$ only depends on the vectors between the two masses $\textbf{r}_1-\textbf{r}_2$, their relative velocity, $\dot{\textbf{r}}_2-\dot{\textbf{r}}_1$, or any higher derivative of $\textbf{r}_1-\textbf{r}_2$. This Newtonian approximation of the problem will be useful in modeling the motion of a binary system during the inspiral phase and before the merger.

    \begin{figure}[t]
    \begin{center}
    \includegraphics[scale=0.5]{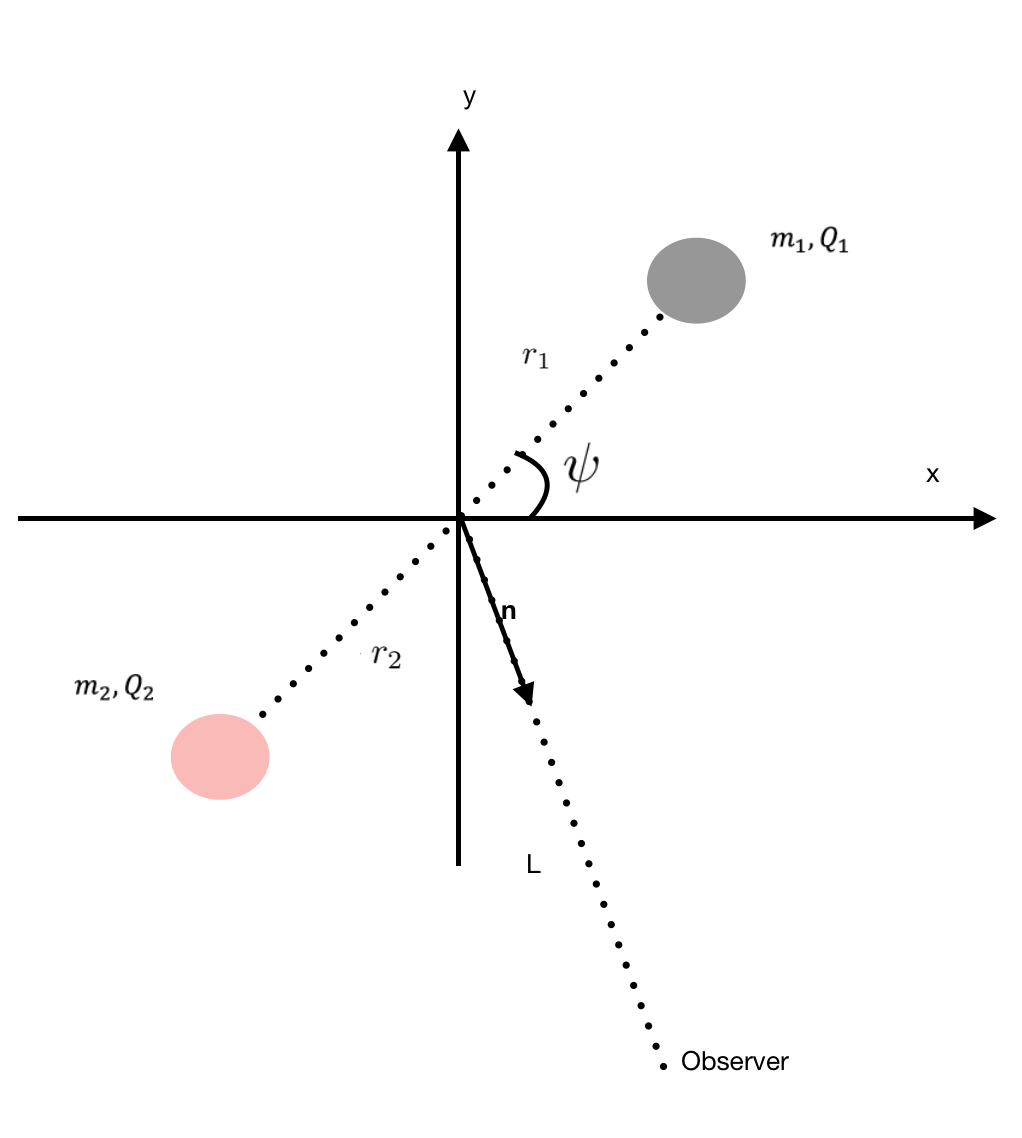}
    \caption{Schematic representation of two point-masses in a Keplerian orbit. Each mass with electric charge $Q_1$ and $Q_2$, respectively. In the figure, $\textbf{n}$ is a unit vector pointing along the observer's direction.\label{fig1a}}
    \end{center}
    \end{figure}

    Let's start by defining the separation between the charged black holes as $\textbf{R}=\textbf{r}_1-\textbf{r}_2$. Therefore, each black hole have coordinates~\cite{Peters:1963ux,Liu:2020cds}
    \begin{equation}
    \label{s2e1}
    \begin{array}{ccc}
    \mathbf{r}_1=r_1(\cos\psi,\sin\psi),&\text{and}&
    \mathbf{r}_2=-r_2(\cos\psi,\sin\psi).
    \end{array}
    \end{equation}
    See the scheme in Fig.~\ref{fig1a}, where $|\mathbf{r_1}|=r_1$ and $|\mathbf{r_2}|=r_2$ are measured with respect to the origin of coordinates. The separation between the black holes is $R=|\textbf{R}|$, and the center of mass of the system is defined by 
    \begin{equation}
    \label{s2e2}
    \mathbf{r}_{CM}=\sum^2_{i=1}\frac{\mathbf{r}_im_i}{M},
    \end{equation}
    with $M=m_1+m_2$ the total mass of the binary system. Hence, from Eq.~(\ref{s2e2}), we obtain that 
    \begin{equation}
    \label{s2e3}
    \mathbf{r}_{CM}=\frac{(r_1m_1-r_2m_2)}{M}\left(\cos\psi,\sin\psi\right).
    \end{equation}
    Since we choose the origin of coordinates the point $(0,0)$, it is clear that 
    \begin{equation}
    \label{s2e4}
    \begin{aligned}
    0&=r_1m_1-r_2m_2\\
    R&=r_1+r_2. 
    \end{aligned}
    \end{equation}
    Therefore,
    \begin{equation}
    \label{s2e5}
    \begin{array}{ccc}
    r_1=\frac{m_2}{M}R,&\text{and}&r_2=\frac{m_1}{M}R.
    \end{array}
    \end{equation}
    From the last expressions we can conclude that $r_1=r_2$ if $m_1=m_2$.
	
	The Lagrangian of the system is given by 
    \begin{equation}
    \label{s2e6}
    \mathcal{L}=\mathcal{T}-\mathcal{U},
    \end{equation}
    where $\mathcal{T}$ and $\mathcal{U}$ denote kinetic and potential energies, respectively. Moreover, $\mathcal{U}$ has two contributions: the gravitational and electric potentials. Therefore, we have
    \begin{equation}
    \label{s2e7}
    \mathcal{U}=-\frac{Gm_1m_2}{R}+\frac{k_e Q_1Q_2}{R}.
    \end{equation}

    On the other hand, from Fig.~\ref{fig1a}, the kinetic energy is given by
    \begin{equation}
    \label{s2e8}
    \mathcal{T}=\frac{1}{2}m_1\dot{r}^2_1+\frac{1}{2}m_2\dot{r}^2_2.
    \end{equation}
    Using Eq.~(\ref{s2e5}), the kinetic energy can be expressed as
    \begin{equation}
    \label{s2e9}
    \mathcal{T}=\frac{1}{2}\mu \dot{R}^2.
    \end{equation}
    Here $\mu$ is known as the \textit{reduced mass} and is defined by the expression
    \begin{equation}
    \label{s2e10}
    \mu=\frac{m_1m_2}{M}.
    \end{equation}
    This means that the problem can be reduced to the motion of a particle with reduce mass $\mu$ and radius $R$. Hence, using polar coordinates $(r,\psi)$, the Lagrangian in Eq.~(\ref{s2e6}) takes the form
    \begin{equation}
    \label{s2e11}
    \mathcal{L}=\frac{1}{2}\mu(\dot{R}^2+R^2\dot{\psi}^2)-\mathcal{U}(R).
    \end{equation}
    Since the Lagrangian only depends on the radial separation, $R$, the system has spherical symmetry. Therefore, the solution is invariant under rotations about any fixed axis. This particular property results in a considerable simplification of the problem because the system's angular momentum is a constant of motion. In addition, from the geometrical point of view, it follows that $\textbf{R}$ is always perpendicular to the fixed direction of the angular momentum $\textbf{L}$, implying that the central force motion always takes place on a plane perpendicular to the polar axis. In this sense, by choosing the polar axis to be in the direction of the angular momentum, we can constrain the discussion to the equatorial plane $\theta=\pi/2$. To understand that conclusion, we can consider the canonical momentum $P_\psi$, given by the following relation
    \begin{equation}
    \label{s2e12}
    P_\psi=\frac{\partial \mathcal{L}}{\partial \dot{\psi}}=\mu R^2\dot{\psi}.
    \end{equation}
    Then, after using the Euler-Lagrange equation\footnote{The Euler-Lagrange equation is given by
    \begin{equation}
    \frac{d}{dt}\left(\frac{\partial\mathcal{L}}{\partial\dot{x}^i}\right)-\frac{d\mathcal{L}}{dx^i}=0,
    \end{equation}
    were $x^i=\psi$ or $R$.},
    we obtain that 
    \begin{equation}
    \label{s2e14}
    \dot{P}_\psi=\frac{d}{dt}\left(\mu R^2 \dot{\psi}\right)=0.
    \end{equation}
    According to Eq.~(\ref{s2e14}), the canonical momentum is a constant of motion known as the angular momentum of the system. Hence,
    \begin{equation}
    \label{s2e15}
    P_\psi=L=\mu R^2\dot{\psi}.
    \end{equation}
    Note that the well-known Kepler's second law can be obtained from Eq.~(\ref{s2e14}) by considering the areal velocity~\cite{Goldstein:1980} 
    \begin{equation}
    \label{s2e17}
    \frac{dA}{dt}=\frac{1}{2}R^2\dot{\psi},
    \end{equation}
    from which 
    \begin{equation}
    \label{s2e18}
    \frac{d}{dt}\left(\frac{dA}{dt}\right)=\frac{d}{dt}\left(\frac{1}{2}R^2\dot{\psi}\right)=0.
    \end{equation}
    Therefore, the conservation of angular momentum is thus equivalent to the constancy of the areal velocity. In other words: ``\textit{The radius vector sweeps out equal areas in equal times}''~\cite{Goldstein:1980}.
	
	On the other hand, after considering the Euler-Lagrange equation for the radial separation and taking into account the conservation of the angular momentum, we can obtain the differential equation 
    \begin{equation}
    \label{s2e22}
    \mu \ddot{R}+\frac{d}{dR}\left(\mathcal{U}+\frac{1}{2}\frac{L^2}{\mu R^2}\right)=0,
    \end{equation}
    which can be expressed in a more suitable way if we multiply the expression by $\dot{R}$. Equation (\ref{s2e22}) then reduces to 
    \begin{equation}
    \label{s2e23}
    \frac{d}{dt}\left(\frac{1}{2}\mu\dot{R}^2+\frac{1}{2}\frac{L^2}{\mu R^2}+\mathcal{U}\right)=0.
    \end{equation}
    Expressed in this way, we can easily see that the quantity inside the brackets is a constant of motion: the well-known energy of the system
    \begin{equation}
    \label{s2e24}
    E=\frac{1}{2}\mu\dot{R}^2+\frac{1}{2}\frac{L^2}{\mu R^2}+\mathcal{U}.
    \end{equation}
	
	On the other hand, according to the Virial theorem, if the forces are derivable from a potential, the kinetic energy and the potential energy are related by the following relation 
	\begin{equation}
	\label{s2e25}
	\mathcal{T}=-\frac{1}{2}\mathcal{U}.
	\end{equation}
	This relation is valid when the forces follow the inverse square law, such as the gravitational and electric forces. Therefore, we can express the total energy as  
	\begin{equation}
	\label{s2e26}
	E=\mathcal{T}+\mathcal{U}=\frac{1}{2}\mathcal{U}=-\frac{Gm_1m_2}{2R}+k_e\frac{ Q_1Q_2}{2R},
	\end{equation}
	where $G$ is the Newton's constant and $k_e$ the electric constant\footnote{Recall that $k_e=1$ in CGS units.}. From Eq.~(\ref{s2e23}), we know that the energy is constant at any value of $R$. In particular, when $R=a$, with $a$ representing the semi-major axis of the Keplerian orbit.  Thus, the energy of the orbit reduces to~\cite{Liu:2020cds}
	\begin{equation}
	\label{s2e27}
	E_{\text{Orbit}}=-\frac{Gm_1m_2}{2a}+k_e\frac{Q_1Q_2}{2a}=-\frac{Gm_1m_2}{2a}(1-\lambda),
	\end{equation}
	where
	\begin{equation}
	\label{s2e28}
	\lambda=k_e\frac{Q_1Q_2}{Gm_1m_2}=\frac{k_e\lambda_1\lambda_2}{G}.
	\end{equation}
    Since the point masses conform a bound system, we have that $\lambda<1$~\cite{Liu:2020cds}.
    
    We can obtain the system's trajectory by solving $R$ and $\dot{\psi}$, with the radial separation expressed as a function of $\psi$ i. e. $R(\psi(t))$. In this sense, it is necessary to obtain a relation between $d/dt$ and $d/d\psi$. To do so, let's consider a function $f(\psi(t))$. Its first derivative takes the form
    \begin{equation}
    \label{s2e29}
    \frac{df}{dt}=\frac{df}{d\psi}\dot{\psi}.
    \end{equation}
    Then, from the conservation of the angular momentum in Eq.~(\ref{s2e15}), we obtain the following relation 
    \begin{equation}
    \label{s2e30}
    \frac{d}{dt}=\frac{L}{\mu R^2}\frac{d}{d\psi}.
    \end{equation}
    In a similar way, the second derivative reduces to
    \begin{equation}
    \label{s2e31}
    \frac{d^2}{dt^2}=\frac{L}{\mu R^2}\frac{d}{d\psi}\left(\frac{L}{\mu R^2}\frac{d}{d\psi}\right).
    \end{equation}
    Therefore, Eq.~(\ref{s2e22}) takes the form 
    \begin{equation}
    \label{s2e32}
    \frac{L}{R^2}\frac{d}{d\psi}\left(\frac{L}{\mu R^2}\frac{dR}{d\psi}\right)-\frac{L^2}{\mu R^3}=-\frac{\partial \mathcal{U}}{\partial R}.
    \end{equation}
    Then, after changing the variable to $u=1/R$, we obtain the following equation~\cite{Goldstein:1980} 
    \begin{equation}
    \label{s2e34}
    \frac{d^2u}{d\psi^2}+u=\frac{\mu\kappa}{L^2},
    \end{equation}
    where we used Eq.~(\ref{s2e7}) and defined  $\kappa\equiv Gm_1m_2(1-\lambda)$. Finally, by doing the change of variable $y=u-\mu\kappa/L^2$, Eq.~(\ref{s2e34}) takes the form~\cite{Goldstein:1980}
    \begin{equation}
    \label{s2e35}
    \frac{d^2y}{d\psi^2}+y=0.
    \end{equation}
    The solution of Eq.~(\ref{s2e35}) has the form $y=C\cos(\psi-\psi_0)$, with $C$ and $\psi_0\equiv\psi(0)$  constants of integration. Therefore, after returning to the original variable $R$, we obtain the following expression~\cite{Goldstein:1980} 
    \begin{equation}
    \label{s2e37}
    R=\frac{L^2}{\mu\kappa[1+\epsilon\cos(\psi-\psi_0)]},
    \end{equation}
    where the eccentricity $\epsilon$ is defined by (see Appendix.~\ref{A1})
    \begin{equation}
    \label{s2e38}
    \epsilon=\frac{CL^2}{\mu\kappa}=\sqrt{1+\frac{2EL^2}{\mu\kappa^2}}.
    \end{equation}
    Note that the factor $L^2/(\mu\kappa)$ in Eq.~(\ref{s2e37}) can be expressed in terms of the eccentricity using Eqs.~(\ref{s2e38}) and (\ref{s2e27}). We obtain~\cite{Goldstein:1980} 
    \begin{equation}
    \label{s2e39}
    \frac{L^2}{\mu\kappa}=-\frac{\kappa(1-\epsilon^2)}{2E}=a(1-\epsilon^2).
    \end{equation}
    As a consequence, the radial separation $R$ can be expressed in terms of the orbital parameters $a$, $\epsilon$ and $\psi$. Hence, Eq.~(\ref{s2e37}) reduces to 
    \begin{equation}
    \label{s2e40}
    R=\frac{a(1-\epsilon^2)}{[1+\epsilon \cos(\psi-\psi_0)]}.
    \end{equation}
    Finally, using Eq~(\ref{s2e38}), we have that
    \begin{equation}
    \label{s2e41}
    \frac{L}{\mu}=\sqrt{G(m_1+m_2)a(1-\epsilon^2)(1-\lambda)}.
    \end{equation}
    Therefore, Eq.~(\ref{s2e15}) takes the form
    \begin{equation}
    \label{s2e42}
    \dot{\psi}=\frac{\sqrt{G(m_1+m_2)a(1-\epsilon^2)(1-\lambda)}}{R^2}.
    \end{equation}
    Equations~(\ref{s2e40}) and (\ref{s2e42}) give the values of $R$ and $\dot{\psi}$ for particles in a Keplerian orbit. Moreover, depending on the eccentricity, one could obtain different trajectories. For example, the orbit would be a circle if $\epsilon =0$, an ellipse if $0<\epsilon<1$, a parabola if $\epsilon =1$, and a hyperbola if $\epsilon>1$.  In this work, we want to investigate the EM radiation of binary systems formed by charged black holes during the inspiral phase. Therefore, we focus on cases where the trajectory is a circular orbit ($\epsilon =0$).
    

    \section{Gravitational and electromagnetic radiation \label{SecIII}}

	In this section, we focus our attention on gravitational and electromagnetic radiations. The mathematical expressions were obtained for the non-charge and charge binary systems in Refs.~\cite{Peters:1963ux}, and, \cite {Liu:2020cds}, respectively. Here, we review and discuss the most crucial aspects. 

    \subsection{Gravitational radiation}
    We start by considering first the gravitational radiation. According to Ref.~\cite{Peters:1963ux}, the total radiation (over all directions of emission) is given by the formula
    \begin{equation}
    \label{s3ae1}
    P_{GW}=\frac{G}{5c^5}\left(\dddot{M}_{ij}\dddot{M}_{ij}-\frac{1}{3}\dddot{M}_{ii}\dddot{M}_{jj}\right).
    \end{equation}
    Here the dot denotes the time derivative and $M_{ij}$ is the mass moment. In the reference frame of an orbit laying on the $(x,y)$ plane (the equatorial plane), $M_{ij}$ takes the form~\cite{Maggiore:2007ulw}
    \begin{equation}
    \label{s3ae2}
    M_{ij}=\mu R
    \left(
    \begin{array}{cc}
    \cos^2\psi&\sin\psi\cos\psi\\
    \sin\psi\cos\psi&\sin^2\psi
    \end{array}
    \right),
    \end{equation}
    where $\mu$ and $R$ are the \textit{reduced mass} and the radial separation between the charged black holes, respectively. Therefore, after using Eq.~(\ref{s2e40}), we obtain \footnote{From now on we set $\psi_0=0$.} 
    \begin{equation}
    \label{s3ae3}
    \begin{aligned}
    M_{11}&=\frac{\mu a^2(1-\epsilon^2)^2}{[1+\epsilon\cos\psi]^2}\cos^2\psi,\\
    M_{12}&=M_{21}=\frac{\mu a^2(1-\epsilon^2)^2}{[1+\epsilon\cos\psi]^2}\sin\psi\cos\psi,\\
    M_{22}&=\frac{\mu a^2(1-\epsilon^2)^2}{[1+\epsilon\cos\psi]^2}\sin^2\psi.
    \end{aligned}
    \end{equation}
	
	According to Eq.~(\ref{s3ae1}), to obtain the total radiation $P_{GW}$ it is necessary to compute the third derivative of $M_{ij}$. Nevertheless, since the components $M_{ij}$ depend on $\psi$, the easiest way to compute their derivatives is using Eq.~(\ref{s2e42}), which can be expressed as
    \begin{equation}
    \label{s3ae4}
    \dot{\psi}=\omega_s=\sqrt{\frac{G(m_1+m_2)(1-\lambda)}{a^3}}(1-\epsilon^2)^{-\frac{3}{2}}(1+\epsilon\cos\psi)^2.
    \end{equation}
    Therefore, after using the chain rule, $\dot{M}_{ij}$ in terms of $\dot{\psi}$ is given by the expression
    \begin{equation}
    \label{s3ae5}
    \dot{M}_{ij}=\frac{dM_{ij}}{d\psi}\dot{\psi}.
    \end{equation}
    The same idea can be extended to the second and third derivatives. Then, we obtain  
    \begin{equation}
    \label{s3ae7}
    \begin{aligned}	\dddot{M}_{11}&=\beta(1+\epsilon\cos\psi)^2[2\sin2\psi+3\epsilon\sin\psi\cos^2\psi],\\
    \dddot{M}_{22}&=\beta(1+\epsilon\cos\psi)^2[-2\sin2\psi-\epsilon\sin\psi(1+3\cos^2\psi)],\\	\dddot{M}_{12}&=\beta(1+\epsilon\cos\psi)^2[-2\cos2\psi+\epsilon\cos\psi(1-3\cos^2\psi)],
    \end{aligned}
    \end{equation}
    where we define  
    \begin{equation}
    \label{s3ae8}
    \beta=\frac{2\mu[G(m_1+m_2)(1-\lambda)]^\frac{3}{2}}{[a(1-\epsilon^2)]^\frac{5}{2}}.
    \end{equation}
    Note that Eq.~(\ref{s3ae7}) reduces to Eqs.~(4.68)-(4.70) reported in Ref.~\cite{Maggiore:2007ulw} when $\lambda=0$, i. e. when  $Q_1=Q_2=0$. Now, from Eq.~(\ref{s3ae1}), the total gravitational radiation is 
    \begin{equation}
    \label{s3ae9}
    P_{GW}=\frac{G}{5c^5}\left[\dddot{M}^2_{11}+\dddot{M}^2_{22}+2\dddot{M}_{12}-\frac{1}{3}(\dddot{M}_{11}+\dddot{M}_{22})^2\right].
    \end{equation}
    Hence, after using Eq.~(\ref{s3ae7}), one obtains~\cite{Liu:2020cds}
    \begin{equation}
    \label{s3ae10}
    \begin{aligned}
    P_{GW}&=\frac{8G^4(m_1+m_2)^3(1-\lambda)^3\mu^2}{15a^5c^5(1-\epsilon^2)^5}(1+\epsilon\cos\psi)^4\\
    &[12(1+\epsilon\cos\psi)^2+\epsilon^2\sin^2\psi],
    \end{aligned}
    \end{equation}
    which reduces to that of Refs.~\cite{Peters:1963ux,Maggiore:2007ulw} when $\lambda=0$.
    
	Usually, the energy of GWs is well defined by considering a temporal average over several periods of a wave~\cite{Maggiore:2007ulw}. This can be done by computing the time average integral
	\begin{equation}
	\label{s3ae11}
	\overline{P}_{GW}=\frac{1}{T}\int^{2\pi}_0\frac{P_{GW}(\psi)}{\dot{\psi}}d\psi.
	\end{equation}
	After using Eqs.~(\ref{s2e42}) and (\ref{s3ae10}), the last expression reduces to
    \begin{equation}
    \label{s3ae12}
    \begin{aligned}
    \overline{P}_{GW}&=\frac{8G^4(m_1+m_2)^3(1-\lambda)^3\mu^2}{15a^5c^5}(1-\epsilon^2)^{-\frac{7}{2}}\\
    &\int^{2\pi}_0\frac{1}{2\pi}[12(1+\epsilon\cos\psi)^4+\epsilon^2\sin^2\psi(1+\epsilon\cos\psi)^2]d\psi,
    \end{aligned}
    \end{equation}
	where we had into account that~\cite{Liu:2020cds} 
	\begin{equation}
	\label{s3ae13}
	T=2\pi\sqrt{\frac{a^3}{G(m_1+m_2)(1-\lambda)}}.
	\end{equation}
	After integration, we obtain 
	\begin{equation}
	\label{s3ae14}
	\overline{P}_{GW}=\frac{32G^4(m_1+m_2)^3(1-\lambda)^3\mu^2}{5a^5c^5(1-\epsilon^2)^{\frac{7}{2}}}\left(1+\frac{73}{24}\epsilon^2+\frac{37}{96}\epsilon^4\right),
	\end{equation}
	which reduces to the expression obtained by P.~C.~Peters and J.~Mathews in Ref.~\cite{Peters:1963ux} when $\lambda=0$. Finally, the average energy loss over an orbital period $T$ due to gravitational radiation is given by
	\begin{equation}
	\label{s3ae15}
	\overline{\frac{dE_{GW}}{dt}}=-\overline{P}_{GW}.
	\end{equation}
	
	
	\subsection{Electromagnetic radiation}
	The rate of emission due to the electromagnetic radiation is given by~\cite{Landau:1975pou}
	\begin{equation}
	\label{s3be1}
	\frac{dE_{EM}}{dt}=-\frac{2\ddot{p}^2}{3c^3}
	\end{equation}
	where $p=|\textbf{p}|$ is the electric dipole moment, which is defined by
	\begin{equation}
	\label{s3be2}
	\textbf{p}=\sum^2_{i=1}Q_i\textbf{r}_i.
	\end{equation}
    In our case, the electric dipole moment of a binary system formed by charged black holes is 
    \begin{equation}
    \label{s3be3}
    \textbf{p}=Q_1\textbf{r}_1+Q_2\textbf{r}_2.
    \end{equation}
    Now, from Eqs.~(\ref{s2e1}) and (\ref{s2e5}), the electric dipole moment reduces to 
    \begin{equation}
    \label{s3eb4}
    \textbf{p}=\frac{Q_1m_2-Q_2m_1}{m_1+m_2}R(\cos\psi,\sin\psi),
    \end{equation}
    from which, after taking into account Eqs.~(\ref{s2e40}) and (\ref{s3ae4}), we obtain  
    \begin{equation}
    \label{s3be5}
    \ddot{\textbf{p}}=-\frac{G(Q_1m_2-Q_2m_1)(1-\lambda)}{a^2(1-\epsilon^2)^2}(1+\epsilon\cos\psi)^2(\cos\psi,\sin\psi).
    \end{equation}
    Hence, $\ddot{p}^2$ is given by 
    \begin{equation}
    \label{s3be6}
    \ddot{p}^2=\ddot{\textbf{p}}\cdot\ddot{\textbf{p}}=\frac{G^2(Q_1m_2-Q_2m_1)^2(1-\lambda)^2(1+\epsilon\cos\psi)^4}{a^4(1-\epsilon^2)^4},
    \end{equation}
    and
    \begin{equation}
    \label{s3be7}
    \frac{dE_{EM}}{dt}=-\frac{2G^2(Q_1m_2-Q_2m_1)^2(1-\lambda)^2(1+\epsilon\cos\psi)^4}{3c^3 a^4(1-\epsilon^2)^4}
    \end{equation}
    Since we are interested in the average energy loss over an orbital period $T$, it is necessary to compute the integral
    \begin{equation}
    \label{s3be8}
    \overline{\frac{dE_{EM}}{dt}}=\frac{1}{T}\int^{2\pi}_0\frac{dE_{EM}}{dt}\dot{\psi}^{-1}d\psi.
    \end{equation}
    Whit the help of Eqs.~(\ref{s3ae4}) and (\ref{s3ae13}), the last integral reduces to 
    \begin{equation}
    \label{s3be9}
    \begin{aligned}
    -\int^{2\pi}_0\frac{G^2(Q_1m_2-Q_2m_1)^2(1-\lambda)^2(1+\epsilon\cos\psi)^2}{3\pi c^3a^4(1-\epsilon^2)^\frac{5}{2}}d\psi,
    \end{aligned}
    \end{equation}
    from which~\cite{Liu:2020cds} 
    \begin{equation}
    \label{s3be10}
    \overline{\frac{dE_{EM}}{dt}}=-\frac{G^2(Q_1m_2-Q_2m_1)^2(2+\epsilon^2)(1-\lambda)^2}{3c^3 a^4(1-\epsilon^2)^\frac{5}{2}}.
    \end{equation}
    \section{Evolution of the orbital parameters\label{SecIV}}
    
    From the physical point of view, a binary system in a Keplerian motion radiates energy and angular momentum. On the other hand, under the approximation of point-like bodies without an intrinsic spin, those quantities are drained from the orbital motion. In this sense, the orbit experiences changes in its semi-major axis and eccentricity until the system reaches the merging phase and collapses. As shown in Ref.~\cite{Liu:2020cds}, the emission of angular momentum has two contributions. The first one is due to the gravitational interaction of the masses. The second one comes as a consequence of the electric interaction of the charges. In this section, we review Refs.~\cite{Maggiore:2007ulw, Liu:2020cds} to compute the evolution of the Keplerian orbit as the binary system realizes energy and angular momentum.   
    
    In the quadrupole approximation, the angular momentum radiated by an orbit on the equatorial plane (see Fig.~\ref{fig1a}) is given   by~\cite{Maggiore:2007ulw,Liu:2020cds} 
    \begin{equation}
    \label{s4e1}
    \overline{\frac{dL_{GW}}{dt}}=\frac{4G}{5c^5}\frac{1}{T}\int^{2\pi}_0\frac{\ddot{M}_{12}(\dddot{M}_{11}-\dddot{M}_{22})}{\dot{\psi}}d\psi.
    \end{equation}
    Hence, from Eq.~(\ref{s3ae3}) and the chain rule, we obtain
    \begin{equation}
    \label{s4e2}
    \begin{aligned}
    \ddot{M}_{12}&=\dot{\psi}\frac{d}{d\psi}\left(\dot{\psi}\frac{dM_{12}}{d\psi}\right)=-\frac{G(m_1+m_2)(1-\lambda)\mu}{a(1-\epsilon^2)}\\
    &\times\sin\psi\left[4\cos\psi+\epsilon(3+\cos2\psi)\right].
    \end{aligned}
    \end{equation}
    After integration, the radiation of angular momentum (the average over one period) due to the gravitational interaction is therefore~\cite{Liu:2020cds}
    \begin{equation}
    \label{s4e3}
    \overline{\frac{dL_{GW}}{dt}}=-\frac{32}{5}\frac{G^\frac{7}{2}\mu^2(m_1+m_2)^\frac{5}{2}(1-\lambda)^\frac{5}{2}}{c^5a^\frac{7}{2}(1-\epsilon^2)^2}\left(1+\frac{7}{8}\epsilon^2\right).
    \end{equation}
    
    Now, let's consider the emission of angular momentum due to the electromagnetic interaction. In Ref.~\cite{Landau:1975pou,Liu:2020cds}, it was shown that the rate of angular momentum carried by the electromagnetic waves is given by
    \begin{equation}
    \label{s4e4}
    \frac{dJ_{EM}}{dt}=-\frac{2}{3c^3}\epsilon^{ikl}\dot{p}_k\ddot{p}_l.
    \end{equation}
    Hence, the last expression takes the form 
    \begin{equation}
    \label{s4e5}
    \frac{dJ_{EM}}{dt}=-\frac{2}{3c^3}(\dot{p}_y\ddot{p}_x-\dot{p}_x\ddot{p}_y).
    \end{equation}
    The first derivative of $\textbf{p}$ can be computed with the help of Eq.~(\ref{s3ae4}). One obtains
    \begin{equation}
    \label{s4e6}
    \dot{\textbf{p}}=\frac{G^\frac{1}{2}(m_2Q_1-m_1Q_2)(1-\lambda)^\frac{1}{2}}{(m_1+m_2)^\frac{1}{2}a^\frac{1}{2}(1-\epsilon^2)^\frac{1}{2}}(-\sin\psi,(\epsilon+\cos\psi)).
    \end{equation}
    Therefore, Eq.~(\ref{s4e5}) takes the form 
    \begin{equation}
    \label{s4e7}
    \frac{dJ_{EM}}{dt}=\frac{G^\frac{3}{2}(m_2Q_1-m_1Q_2)^2(1-\lambda)^\frac{3}{2}(1+\epsilon\cos\psi)^3}{6\pi a^\frac{5}{2}(m_1+m_2)^\frac{1}{2}(1-\epsilon^2)^\frac{5}{2}}.
    \end{equation}
    The average emission of angular momentum in one period is given by
    \begin{equation}
    \label{s4e8}
    \overline{\frac{dL_{EM}}{dt}}=\frac{1}{T}\int^{2\pi}_0\frac{dJ_{EM}}{dt}\dot{\psi}^{-1}d\psi.
    \end{equation}
    From which, 
    \begin{equation}
    \label{s4e9}
    \begin{aligned}
    \overline{\frac{dL_{EM}}{dt}}&=
    \frac{2G^\frac{3}{2}(m_2Q_1-m_1Q_2)^2(1-\lambda)^\frac{3}{2}}{3 a^\frac{5}{2}c^2(1-\epsilon^2)(m_1+m_2)^\frac{1}{2}}.
    \end{aligned}
    \end{equation}
    
    In Sec.~\ref{SecII}, due to the virial theorem, the total energy of the Keplerian orbit is given by Eq.~(\ref{s2e27}), from which the semi-major axis is given by 
    \begin{equation}
    \label{s4e10}
    a=\frac{G\mu(m_1+m_2)(1-\lambda)}{2|E_\text{Orbit}|}.
    \end{equation}
    Therefore, after computing the derivative with respect to time $t$, we obtain 
    \begin{equation}
    \label{s4e11}
    \frac{da}{dt}=-\frac{G\mu(m_1+m_2)(1-\lambda)}{2}\frac{E_\text{Orbit}}{|E_\text{Orbit}|^3}\frac{dE_\text{Orbit}}{dt}.
    \end{equation}
    According to Ref.~\cite{Peters:1963ux,Maggiore:2007ulw}, in the case of a binary system formed by to non-charged black holes, the emission of GWs costs energy. Therefore, the variation in $E_\text{Orbit}$ must be equal to the power radiated by GWs. Nevertheless, in the case of two charged black holes, the variation in the total energy ($E_\text{Orbit}$) has two contributions: the power radiated by the gravitational wave, $\overline{dE_{GW}/dt}$, and the power radiated by the electromagnetic field, $\overline{dE_{EM}/dt}$. Hence, we have the following relation
    \begin{equation}
    \label{s4e12}
    \frac{dE_{\text{Orbit}}}{dt}=\overline{\frac{dE_{GW}}{dt}}+\overline{\frac{dE_{EM}}{dt}}.
    \end{equation}
    After replacing into Eq.~(\ref{s4e11}), we get the following expression~\cite{Liu:2020cds}
    \begin{equation}
    \label{s4e13}
    \begin{aligned}
    \frac{da}{dt}&=-\frac{2G^3(m_1+m_2)^2(1-\lambda)^2\mu}{15a^3c^5(1-\epsilon^2)^\frac{7}{2}}(96+292\epsilon^2+37\epsilon^4)\\
    &-\frac{2(\epsilon^2+2)G(m_1+m_2)(1-\lambda)(\lambda_1-\lambda_2)^2\mu}{3c^3a^2(1-\epsilon)^\frac{5}{2}}.
    \end{aligned}
    \end{equation}
    On the other hand, from Eq.~(\ref{s2e41}), we have that 
    \begin{equation}
    \label{s4e14}
    \frac{dL_\text{Orbit}}{dt}=\mu\sqrt{G(m_1+m_2)(1-\lambda)}\left[(1-\epsilon^2)\frac{da}{dt}-2a\epsilon\frac{d\epsilon}{dt}\right].
    \end{equation}
    Then, after solving for $d\epsilon/dt$, we obtain
    \begin{equation}
    \label{s4e15}
    \frac{d\epsilon}{dt}=\frac{(1-\epsilon^2)}{2a\epsilon}\frac{da}{dt}-\frac{\sqrt{a(1-\epsilon^2)}}{a\epsilon\mu\sqrt{G(m_1+m_2)(1-\lambda)}}\frac{dL_\text{Orbit}}{dt}.
    \end{equation}
    Using Eq.~(\ref{s4e14}) and taking into account that 
    \begin{equation}
    \label{s4e16}
    \frac{dL_\text{Orbit}}{dt}=\overline{\frac{dL_{GW}}{dt}}+\overline{\frac{dL_\text{EM}}{dt}},
    \end{equation}
    Eq.~(\ref{s4e15}) reduces to~\cite{Liu:2020cds} 
    \begin{equation}
    \label{s4e17}
    \begin{aligned}
    \frac{d\epsilon}{dt}&=-\frac{\epsilon (121\epsilon^2+304)G^3(1-\lambda)^2m_1m_2(m1+m2)}{15a^2c^5(1-\epsilon^2)^\frac{5}{2}}\\
    &-\frac{\epsilon G(1-\lambda)(m_2Q_1-m_1Q_2)}{4\pi a^3(1-\epsilon^2)^\frac{3}{2}m_1m_2}.
    \end{aligned}
    \end{equation}
    The system of differential equations (\ref{s4e13}) and (\ref{s4e17}) describes the evolution due to gravitational and electrical interactions of the binary system. Note that the system is coupled. Therefore, any variation on the eccentricity $\epsilon$ has a repercussion on the evolution of the semi-major axis $a$. In the next section, we discuss the particular case of circular orbits. 
    
    \section{Quasi-circular approximation \label{SecV}}
    
    It is well-known that a binary system circularizes its orbit after some time. In this sense, during the inspiral phase, the system loses energy in a way that the motion of the binary system remains circular. This can be seen from Eq.~(\ref{s4e15}). When the orbit reaches the value $\epsilon=0$ (circular orbit), the binary system will continue moving in a circular orbit because $\dot{\epsilon}=0$. Therefore, if the initial conditions of the binary system are those of a circular orbit, the system will continue its circular motion with a variation of $dR/dt$ given by\footnote{When $\epsilon=0$, the semi-major axis $a$ and the separation $R$ are the same. See Eq.~(\ref{s2e40}).}
    \begin{equation}
    \label{s5e1}
    \begin{aligned}
    \frac{dR}{dt}&=-\frac{64G^3(m_1+m_2)^2(1-\lambda)^2\mu}{5R^3c^5}\\
    &-\frac{4G(m_1+m_2)(1-\lambda)(\lambda_1-\lambda_2)^2\mu}{3 c^3R^2}.
    \end{aligned}
    \end{equation}
	\begin{figure*}[t]
	\begin{center}$
	\begin{array}{ccc}
	\includegraphics[scale=0.3]{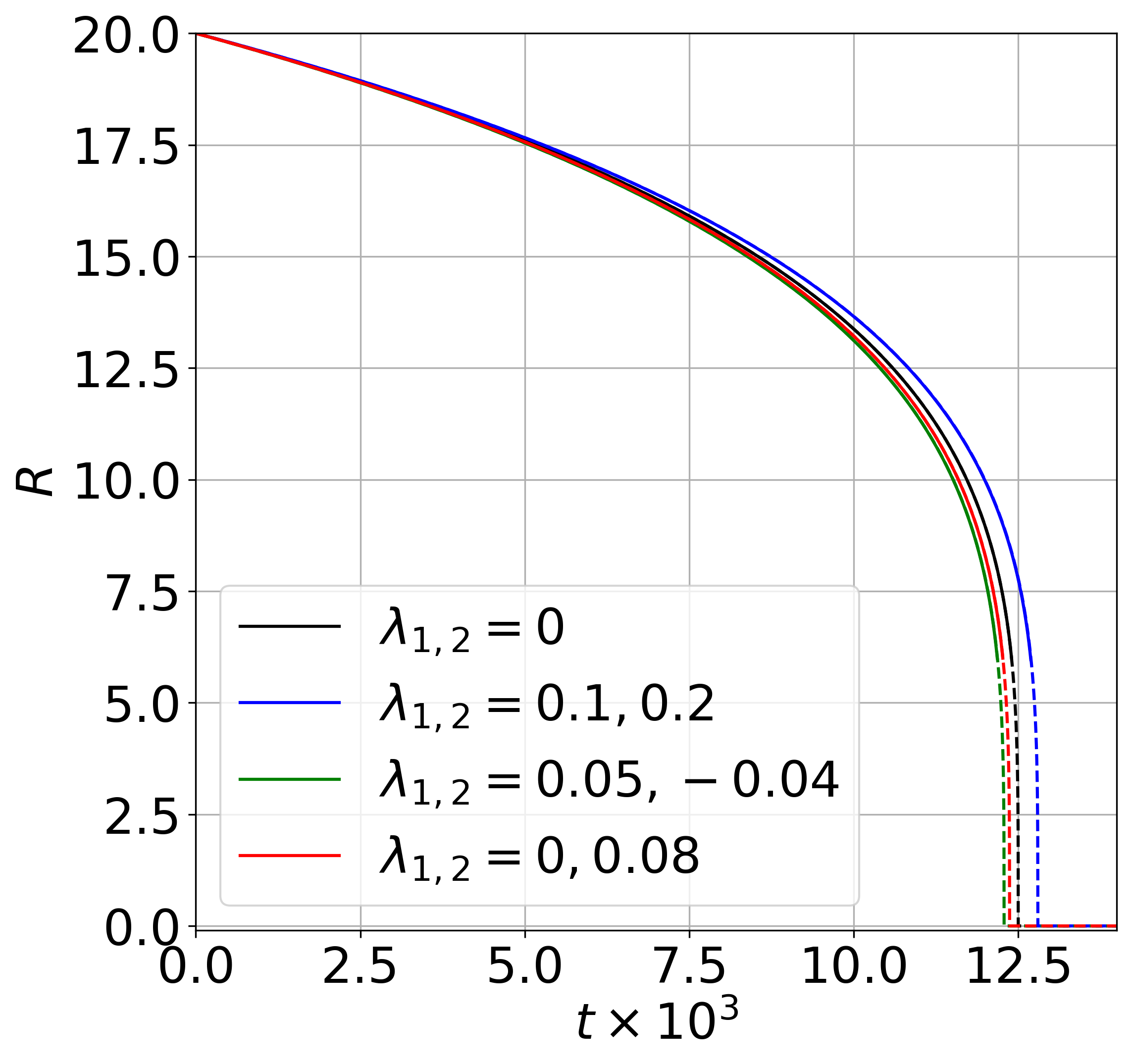}&
	\includegraphics[scale=0.3]{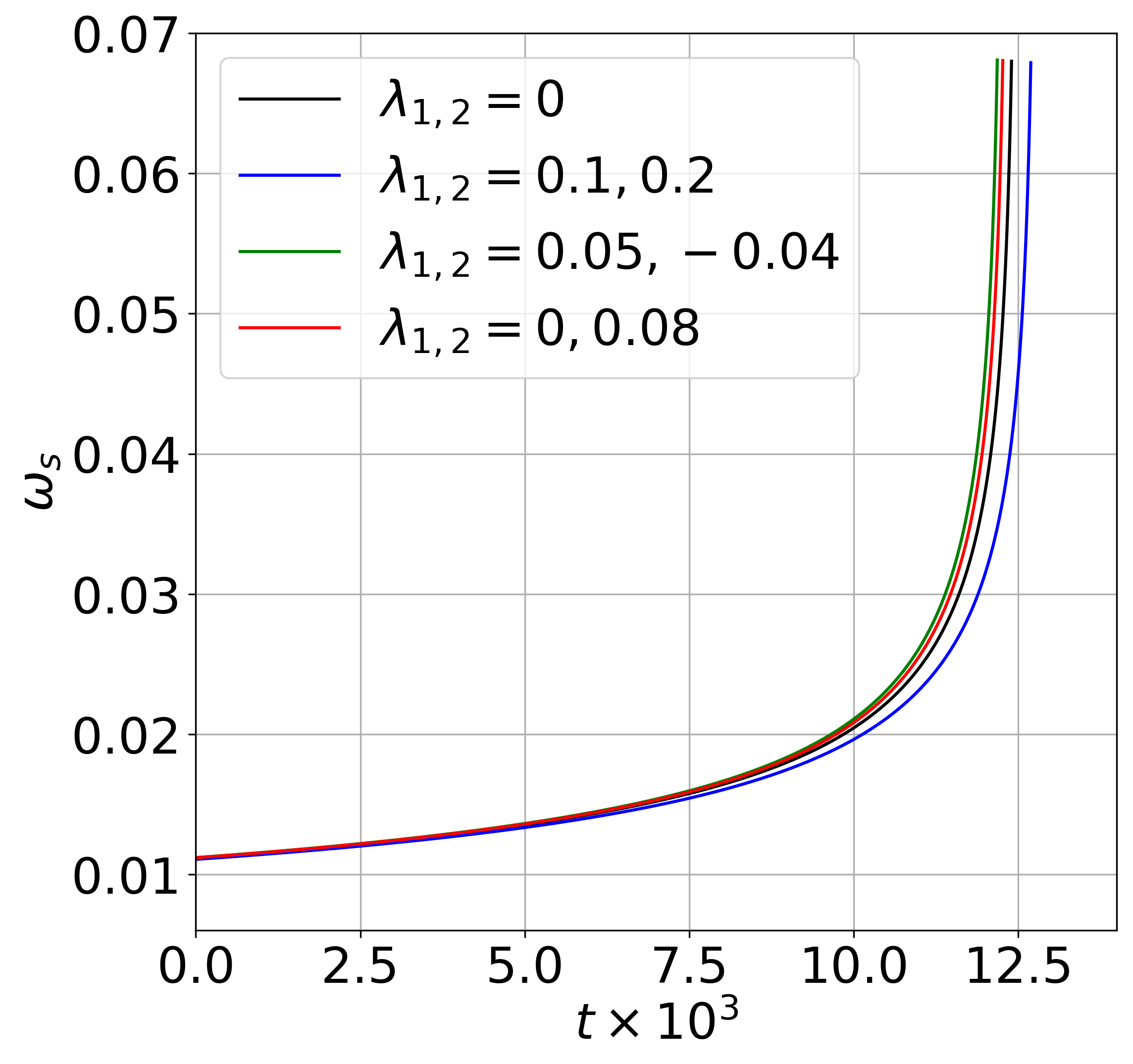}&
	\includegraphics[scale=0.3]{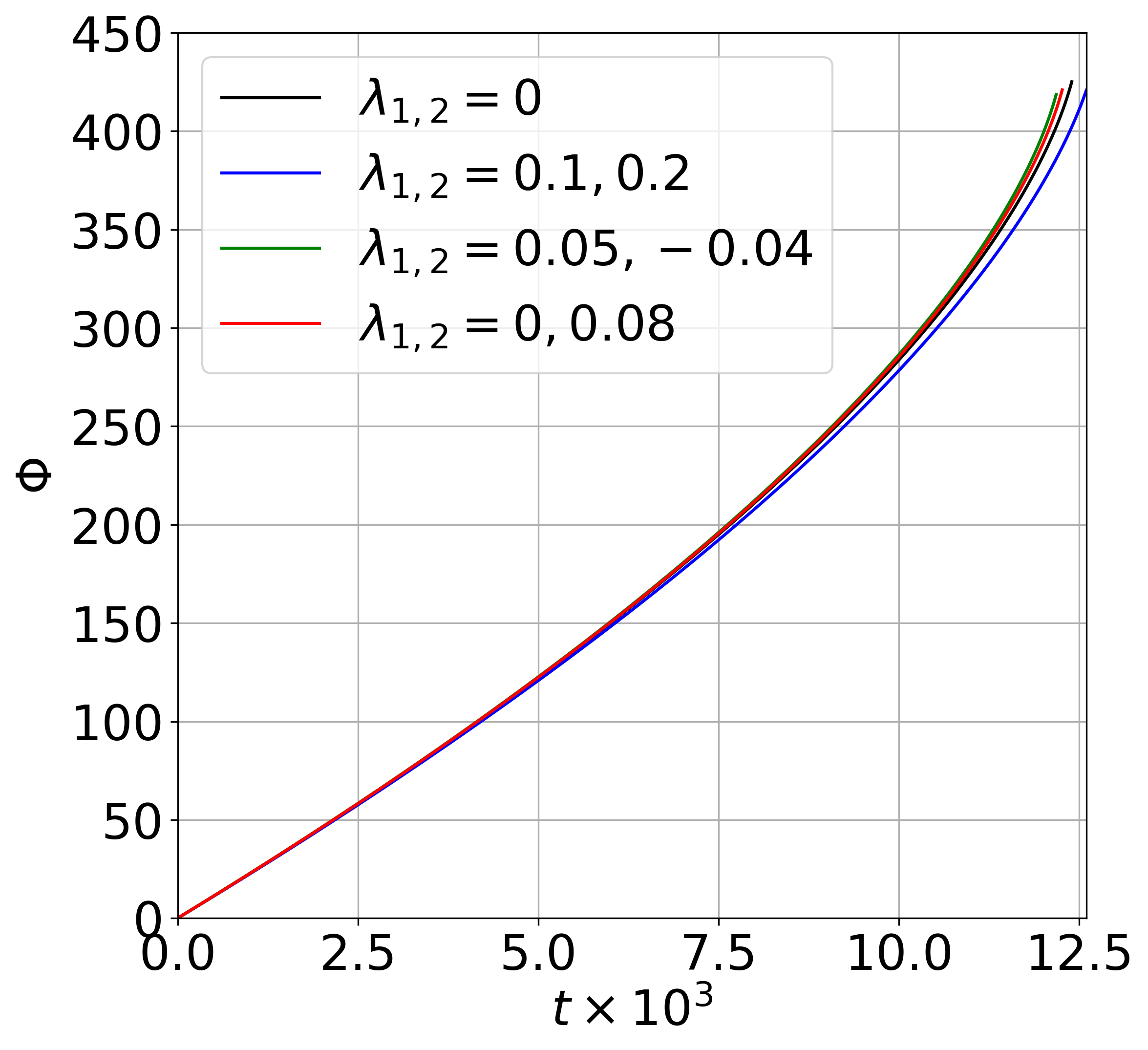}
	\end{array}$
	\end{center}
	\caption{Left panel: $R$ vs. $t$. The dashed lines correspond to the numerical solution of Eq.~(\ref{s5e1}), while the continuous line is the analytical solution. See the first expression in Eq.~(\ref{s5e11}). Central panel: $\omega_s$ vs. $t$, we use the second expression in Eq.~(\ref{s5e11}). Right panel: $\Phi$ vs. $t$, we use the third expression in Eq.~(\ref{s5e11}). For the plot we consider $m_1=m_2=1/2$, $R_0=20$ and $\psi_0=0$.  \label{fig2}}
	\end{figure*}
	\begin{figure*}[t]
	\begin{center}
	\includegraphics[scale=0.18]{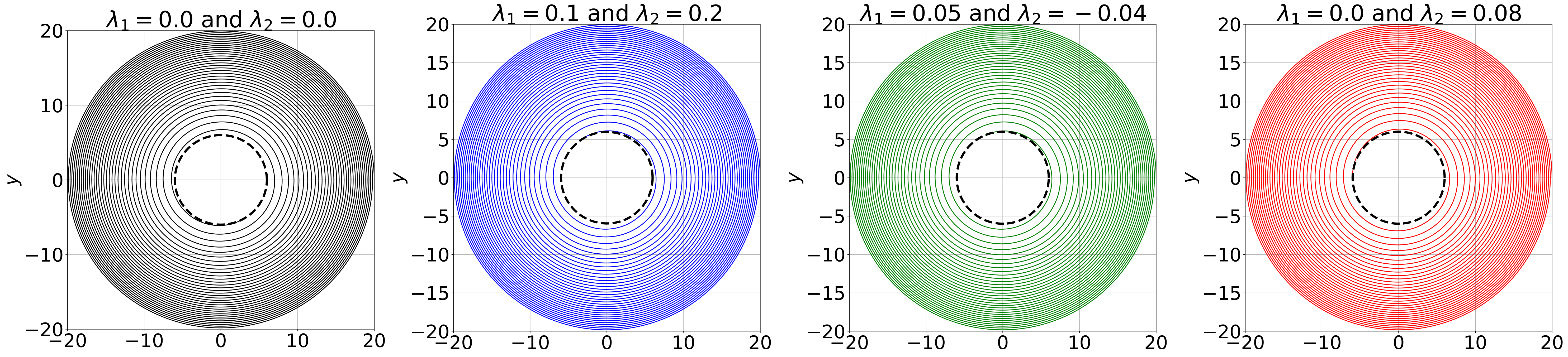}
	\includegraphics[scale=0.18]{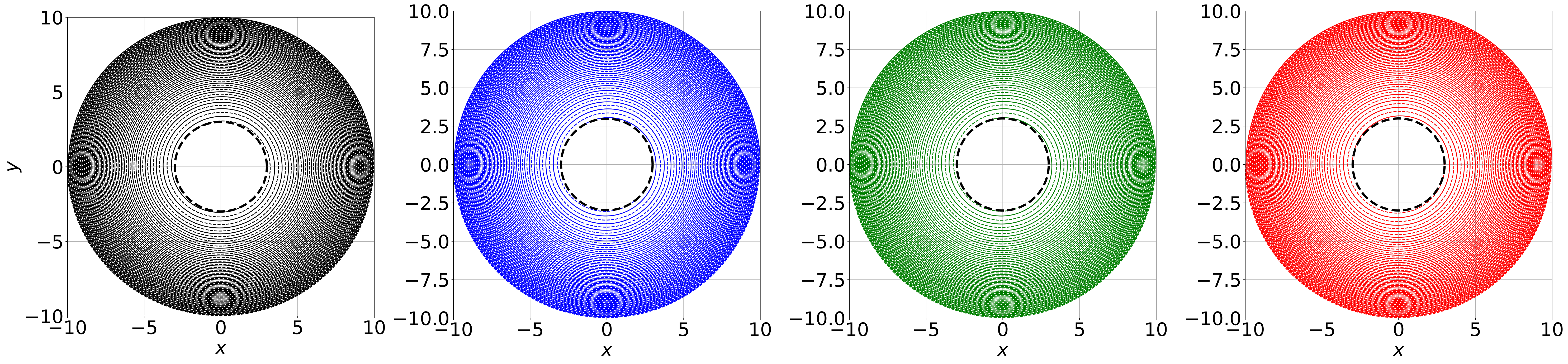}
	\caption{The \textit{reduced mass} (top panel) and black holes' trajectories (button panel) of the binary system for different values of $\lambda_1$ and $\lambda_2$. In the plot we consider dimensionless units where $m_1=m_2=1/2$, $\mu=1/4$, $R_0=20$ and $\psi_0=0$. In the button panel, the continuous line shows the trajectory of the black hole $1$ while the dashed line that of black hole $2$. $R_\text{ISCO}$ is represented by a thick black-dashed circumference. \label{fig3}}
	\end{center}
	\end{figure*}
    From the last equation, we can identify several situations. First, if $\lambda_1=\lambda_2=0$, Eq.~(\ref{s5e1}) reduces to that obtained by M.~Maggiore in Ref.~\cite{Maggiore:2007ulw}. Secondly, if the two black holes carry the same charge-to-mass ratio, $\lambda_1 =\lambda_2$, the second term in the right-hand side of Eq.~(\ref{s5e1}) vanishes. Therefore, the electric dipole vanishes, and it is necessary to consider the next order, whose term decomposes into the charge quadrupole and current dipole that generate electric quadrupolar and magnetic dipolar radiation, respectively, as the leading order contributions. Finally, if the charge-to-mass-ratio difference is small, both the GW quadrupole and the EM dipole emissions can be important~\cite{Christiansen:2020pnv}. In this manuscript, we focus on the dipole order of EM waves. 
    
    Before continuing our discussion, it is important to express Eq.~(\ref{s5e1}) in terms of dimensionless quantities. To do so, we follow Ref.~\cite{Maggiore:2007ulw} and define the dimensionless variables by the relations
    \begin{equation}
    \label{s5e2}
    \begin{array}{ccc}
    R\rightarrow\frac{R}{R_*}&\text{and}&t\rightarrow\frac{ct}{R_*},
    \end{array}
    \end{equation}
    with
    \begin{equation}
    \label{s5e3}
    R^3_*=\left(\frac{2GM}{c^2}\right)^2\left(\frac{G\mu}{c^2}\right).
    \end{equation}
    Thus, Eq.~(\ref{s5e1}) takes the form
    \begin{equation}
    \label{s5e4}
    \frac{dR}{dt}=-\frac{\alpha}{R^3}-\frac{\beta}{R^2},
    \end{equation}
    with $\alpha$ and $\beta$ dimensionless constants, see Appendix~\ref{A0}. 
    
    The solution of Eq.~(\ref{s5e4}), can be expressed as 
    \begin{equation}
    \label{s5e7}
    \alpha t=\int^{R_0}_{R}
    \frac{R'^3dR'}{1+\gamma R'},
    \end{equation}
    where
    \begin{equation}
    \label{s5e8}
    \gamma=\frac{\beta}{\alpha}=\frac{5}{48}\frac{(\lambda_1-\lambda_2)^2}{(1-\lambda)}\left(\frac{4\mu}{M}\right)^{1/3}.
    \end{equation}
    Hence, after integration, the solution can be represented as $\alpha t = f(R_0)-f(R)$, where the function $f(R)$ is defined by~\cite{Christiansen:2020pnv}
    \begin{equation}
    \label{s5e9}
    f(R)=-\frac{\log (1+\gamma R)}{\gamma^4}+\frac{R}{\gamma^3}-\frac{R^2}{2 \gamma^2}+\frac{R^3}{3\gamma}.
    \end{equation}
    One can use the function $f(R)$ to define the time to coalescence $\tau(R)=f(R)/\alpha$, from which
    \begin{equation}
    \label{s5e9a}
    t=\tau_0-\tau(R).
    \end{equation}
    Note that $R=R_0$ when $t=0$. Therefore, $\tau_0$ corresponds to the time at the coalescence $t_\text{coal}$ and Eq.~(\ref{s5e9a}) reduces to well-known definition $\tau=t_\text{coal}-t$~\cite{Maggiore:2007ulw}.
    
    On the other hand, the form of $f(R)$ makes it difficult to investigate the dynamics of the binary system analytically due to the term with the $\log$ function. Nevertheless, by considering a small charge-to-mass-ration, it is possible to obtain a simple expression for $f(R)$ when $\gamma R<<1$~\cite{Christiansen:2020pnv}. In this sense, after expanding the $\log(1+\gamma R)$ up $6^\text{th}$ order, the time to the coalescence $\tau$ takes the following form 
    \begin{equation}
    \label{s5e10}
    \tau(R)\approx\frac{R^4}{4\alpha}\left(1-\frac{4\gamma R}{5}\right).
    \end{equation}
    Note that Eq.~(\ref{s5e10}) reduces to Eq.~(4.26) of Ref.~\cite{Maggiore:2007ulw} when $\gamma=0$. Using this approximation and defining $u=\tau/\tau_0$, it is possible to obtain the following expressions for $R$, $R_0$, $\omega_s$, $\omega_0$ and $\Phi$ (See Appendix~\ref{A2} for details)
    \begin{equation}
    \label{s5e11}
    \begin{aligned}
    \frac{R}{R_0}&=u^{1/4}\left[1-\frac{\gamma R_0}{5}\left(1-u^{1/4}\right)\right],\\\\
    \frac{\omega_s}{\omega_0}&=u^{-3/8}\left[1+\frac{3}{10}\delta\omega^{-2/3}_0(1-u^{1/4})\right],\\\\
    \Phi&=\frac{16\tau_0}{5}\left(\frac{8\tau_0}{3\sigma}\right)^{-3/8}\left[1-u^{5/8}-\frac{3\delta}{14}\left(\frac{8\tau_0}{3\sigma}\right)^{1/4}(1-u^{7/8})\right].
    \end{aligned}
    \end{equation}
    and 
    \begin{equation}
    \label{s5e11a}
    \begin{aligned}
    R_0&=(4\alpha\tau_0)^{1/4}\left[1+\frac{\gamma(4\alpha\tau_0)^{1/4}}{5}\right],\\\\
    \omega_0&=\left(\frac{3\sigma}{8\tau_0}\right)^{3/8}\left[1-\frac{3}{10}\delta\left(\frac{8\tau_0}{3\sigma}\right)^{1/4}\right].
    \end{aligned}
    \end{equation}
    Christiansen et al. obtained similar expressions in Ref.~\cite{Christiansen:2020pnv}, where the charge is not the usual electric charge but a dark sector charge. 
    
    From the Kepler law in Eq.~(\ref{s3ae4}) one can obtain the following relation~\cite{Maggiore:2007ulw}
	\begin{equation}
	    \label{s5e11b}
	    \dot{R}=-\frac{2}{3}R\omega_s\frac{\dot{\omega}_s}{\omega^2_s}.
	\end{equation}
    Note that $|\dot{R}|$ is smaller than the tangential velocity $\omega_s R$ if $\dot{\omega}_s<<\omega^2_s$. Therefore, we can use circular orbits with a slowly varying radius to model the dynamics of the binary system, the well-known quasi-circular approximation. Nevertheless, in the case of the binary system formed by charged black holes, the quasi-circular approximation must include the condition $\gamma R<<1$ so that we can use Eqs.~(\ref{s5e11}) and (\ref{s5e11a}) to model the inspiral phase. 
	\begin{figure*}[t]
		\begin{center}
			\includegraphics[scale=0.35]{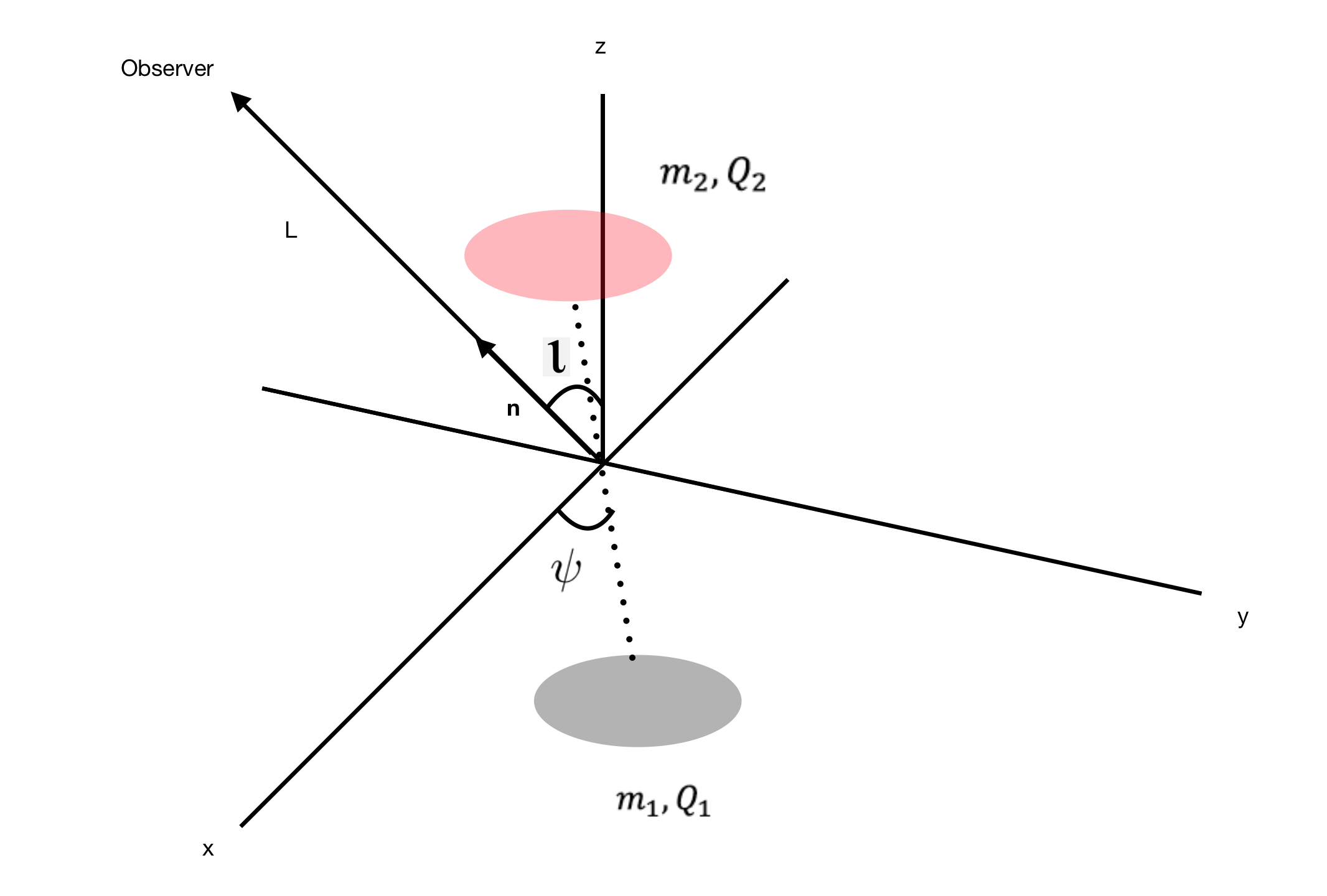}
			\caption{Scheme of two point-masses with electric charge in a Keplerian orbit. In the figure, $\textbf{n}$ is a unit vector pointing in the same direction as the observer, $\iota$ is the angle between the z-axis and $\textbf{n}$, laying on the xz-plane. \label{fig4}}
		\end{center}
	\end{figure*}
	
	In Fig.~\ref{fig2}, we plot the analytical expressions for $R$, $\omega_s$, and $\Phi$ as a function of $t$ for different values of $\lambda_1$ and $\lambda_2$. We also show the numerical solution of Eq.~(\ref{s5e1}). See the dashed lines in the left panel. On the other hand, since the quasi-circular approximation can not describe the motion of the binary system when the radial separation is too small, the analytical solution for $R$ (as well as $\omega_s$ and $\Phi$) is plotted until a certain value, see the continuous lines in the left panel of the figure. In this sense, and following Wang et al., we choose the innermost stable circular orbit (ISCO) as the moment of coalescence~\cite{Wang:2021vmi}. For example, when $\lambda_1=\lambda_2=0$, $R_{ISCO}=6(4\mu/M)^{-1/3}$. Nevertheless, when $\lambda_1=\lambda_2\neq 0$, it is difficult to know exactly the final values of $\lambda$ and $M$ for the remnant black hole. Therefore, we use the following expression (in dimensionless units) to compute $R_{ISCO}$~\cite{Wang:2021vmi} (see Appendix\ref{A0}).
	\begin{equation}
	    \label{s5e12}
	    R_{ISCO}=\frac{4\lambda^2_*}{3+\frac{1}{C}+C}\left(\frac{4\mu}{M}\right)^{-1/3},
	\end{equation}
	where 
	\begin{equation}
	    \label{s5e13}
	    C=-\left(9-8\lambda^2_*-4\sqrt{4\lambda^4_*-9\lambda^2_*+5}\right)^{1/3},
    \end{equation}
    and 
    \begin{equation}
        \label{s5e14}
        \lambda_*=\text{min}\left[\left| \frac{m_1\lambda_1+m_2\lambda_2}{M}\right|,\left| \frac{m_2\lambda_1+m_1\lambda_2}{M}\right|\right].
    \end{equation}
    In table~\ref{table1}, we show some values for $R_{ISCO}$. 
    \begin{table}[t]
    \caption{\label{table1}
    The radial separation at the coalescence, $R_{ISCO}$, for different values of $\lambda_1$ and $\lambda_2$. We consider dimensionless units with $m_1=m_2=1/2$, $M=1$ and $\mu=1/4$.}
    \begin{ruledtabular}
    \begin{tabular}{cccc}
    $\lambda_1$&$\lambda_2$&$\lambda_*$& $R_{ISCO}$\\
    \hline\\
    0.0  & 0.0   & $--$  & 6.0   \\
    0.05 & -0.04 & 0.005 & 5.999  \\
    0.0  & 0.08  & 0.075 & 5.998   \\
    0.1  & 0.2   & 0.15  & 5.966    \\
    \end{tabular}
    \end{ruledtabular}
    \end{table}
	
	The left panel of Fig.~\ref{fig2} shows how the radial separation between the charged black holes reduces in a way that resembles an inspiral. See Fig.~\ref{fig3}, where we plot the \textit{reduced mass} and the black holes' trajectories in the binary system, upper and lower panel, respectively. Note that depending on the values of $\lambda_1$ and $\lambda_2$, the duration of the inspiral phase is longer or shorter. For example, when we consider a binary system formed by positively charged black holes ($\lambda_1=0.1$ and $\lambda_2=0.2$), the inspiral phase is longer than the other three cases. See the continuous blue line in the left panel of Fig.~\ref{fig2}. In the case of two black holes with opposite signs  ($\lambda_1=0.05$ and $\lambda_2=-0.04$), the inspiral phase takes less time than the cases in which the binary system is formed by non-charged black holes or by only one charged black hole ($\lambda_1=0.0$ and $\lambda_2=0.08$), plotted with a continuous red line in the figure. Therefore, the presence of charge in the binary system does affect the coalescence time.
	
    From the physical point of view, this behavior agrees with the phenomenological interaction between electric charges. As shown in Fig.~\ref{fig1a}, we have considered each black hole in the binary system as charged-point masses. In this sense, the electric and gravitational interactions become stronger as the black holes get closer. Nevertheless, the presence of an electric charge produces attraction or repulsion depending on its sign, in contrast to the gravitational interaction, which is always attractive. Hence, in the case of the binary system formed by black holes with opposite signs, the gravitational interaction will encounter an additional attraction that makes the separation $R$ change faster than in the case of a binary system formed by charged black holes with the same sign, where the gravity ``\textit{competes}'' against the repulsion of the electric charges. 
    
    Finally, in the central and left panels of Fig.~\ref{fig2}, we show the behavior of $\omega_s$ and $\Phi$ (respectively) as a function of time $t$. From the figures, it is possible to see how $\omega_s$ increases by time, in agreement with Kepler's law $\gamma R=\delta \omega^{-2/3}_s$, where $\omega_s$ increases as the radial separation $R$ decreases. Moreover, because the inspiral phase takes more time when $\lambda_1=0.1$ and $\lambda_2=0.2$, the value of $\omega_s$ is smaller than in the other cases; see the continuous blue line. On the other hand, when $\lambda_1=0.05$ and $\lambda_2=-0.04$, the inspiral phase is the shortest. Therefore, the values for $\omega_s$ are larger. See the continuous green line. Moreover, note that $\omega_s$ will diverge as $R\rightarrow0$, in contrast to $\Phi$, which has a similar behavior but reaches a finite value when $R\rightarrow0$.
    
    \section{The gravitational and electromagnetic waves\label{SecVI}}
    
    In this section, we will investigate the electromagnetic and gravitational radiation under the approximation $\gamma R<<1$. Therefore, we shall use the expressions obtained in Eq.~(\ref{s5e10}) along with the electrodynamics theory to compute an analytical expression for the electromagnetic field. In this way, and from the phenomenological point of view,  we will be able to understand the electromagnetic counterpart of a binary system formed by charged black holes.
    
    \subsection{The electromagnetic wave}
    
    To compute the EM field generated by the binary system, an observer must reckon that the fields have the retarded value due to the motion of the charges. Hence, if the observer is located at a distance $L$ far from the source (with $L>>R$), the potentials take the form~\cite{Landau:1975pou}
    \begin{equation}
        \label{s5ae1}
        \begin{aligned}
         \varphi&=\frac{1}{L}\int \rho_{t-L/c}dV,\\\\
	     \mathbf{A}&=\frac{1}{cL}\int\mathbf{J}_{t-L/c}dV,
	     \end{aligned}
    \end{equation}
    also known as the Liénard–Wiechert potentials. In the last equation, $\rho$ and $\textbf{J}$ are the charge and current densities evaluated at the retarded time $t_r=t-L/c$, respectively. Note that the potentials in Eq.~(\ref{s5ae1}) reduce to the static case when $\rho$ and $\textbf{J}$ do not depend on time. 
    
    At large distances, the EM field can be considered like a wave plane if one takes small regions of space. Therefore, it is possible to relate the electric and magnetic fields using the following relation~\cite{Landau:1975pou}
    \begin{equation}
        \label{s5ae2}
        \textbf{E}=\textbf{B}\times\textbf{n},
    \end{equation}
    where $\textbf{n}$ is a unit vector in the direction of $L$, see the scheme in Fig.~\ref{fig4}. From the last equation, one concludes that $\textbf{E}$ and $\textbf{B}$ are perpendicular to each other. Therefore, since $\textbf{B}=\nabla\times\textbf{A}$, one only needs to compute the vector potential $\textbf{A}$ for a complete determination of the EM field in the \textit{wave zone}, which is the name of the region where the wave plane approximation takes place. In this zone, the vector potential takes the form~\cite{Landau:1975pou}
    \begin{equation}
        \label{s5ae3}
        \textbf{A}=\frac{\dot{\textbf{p}}}{cL}+\frac{\ddot{\textbf{D}}}{6c^2L}+\frac{\bm{\dot{\mu}}\times\textbf{n}}{cL},
    \end{equation}
    where $\textbf{p}$ is the \textit{dipole moment} of the system defined in Eq.~(\ref{s3be2}). $\textbf{D}$ is the \textit{quadrupole moment} of the system with components\footnote{$D_{\alpha\beta}$ is the quadrupole moment tensor with null trace $D_{\alpha\alpha}=0$.} $D_\alpha= D_{\alpha\beta}n_\beta$~\cite{Landau:1975pou}  
    \begin{equation}
        \label{s5ae4}
        D_{\alpha\beta}=\sum_iQ_i(3x_\alpha x_\beta-\delta_{\alpha\beta} r^2_i),
    \end{equation}
    and $\bm{\mu}$ is the \textit{magnetic moment}, given by the relation~\cite{Landau:1975pou}
    \begin{equation}
    \label{s5ae5}
        \bm{\mu}=\frac{1}{2c}\sum_i Q_i\textbf{r}_i\times \textbf{n}.
    \end{equation}
    In the last expressions, i. e. Eqs.~(\ref{s5ae4}) and (\ref{s5ae5}), the sum goes over all charges, the dot $\dot{{}}$ denotes derivative with respect to time and $x_\alpha$ are the components of $\textbf{r}$ for each charge. Hence, after computing $\nabla\times\textbf{A}$, we have that the EM field is given by the following expression (see Ref.~\cite{Landau:1975pou} for details)
    \begin{equation}
    \label{s5ae6}
        \begin{aligned}
        \textbf{B}&=\frac{1}{c^2 L}\left\{\ddot{\textbf{p}}\times\textbf{n}+\frac{1}{6c}\dddot{\textbf{D}}\times\textbf{n}+(\ddot{\bm{\mu}}\times\textbf{n})\times\textbf{n}\right\},
        \end{aligned}
    \end{equation}
    evaluated at the retarded time. From Eq.~(\ref{s5ae6}), we identify the contributions to the EM field of the dipole (first term), quadrupole (second term), and magnetic moment (third term). In this manuscript, we only consider the dipole contribution in the results, but we compute the quadrupole contribution for completeness. In this sense, we only use the first two terms of Eq.~(\ref{s5ae6}).
    
    According to Fig.~\ref{fig4}, the observer is located at a distance $L$ along the direction of the unit vector $\textbf{n}$, which forms an angle $\iota$ with the z-axis. Hence, 
    \begin{equation}
        \label{s5e7}
        \textbf{n}=(\sin\iota,0,\cos\iota),
    \end{equation}
    and, from the first term in Eq.~(\ref{s5ae6}) and Eq.~(\ref{s3be5}), the dipole contribution is given by 
    \begin{equation}
        \label{s5e7a}
        \textbf{B}_\text{dipole}=\frac{1}{c^2 L}\ddot{\textbf{p}}\times\textbf{n}.
    \end{equation}
    In dimensionless units, the last expression reduces to
     \begin{equation}
        \label{s5e8}
        \begin{aligned}
        \textbf{B}_\text{dipole}&=\frac{m_1m_2(\lambda_1-\lambda_2)(1-\lambda)}{ R^2L}\\
        &\times (-\sin\psi\cos\iota,\cos\psi\cos\iota,\sin\psi\sin\iota),
        \end{aligned}
     \end{equation}
    see Appendix~\ref{A0}. Note that for $\lambda_1=\lambda_2$, the dipole contribution vanishes and it is necessary to consider the quadrupole contribution.  
    
    The quadrupole radiation is given by 
    \begin{equation}
        \label{s5ae9}
        \textbf{B}_{\text{quadrupole}}=\frac{1}{6c^3L}\dddot{\textbf{D}}\times\textbf{n}.
    \end{equation}
   As mentioned before, the vector $\textbf{D}$ can be computed as the projection of the tensor $D_{\alpha\beta}$ along the unit vector $\textbf{n}$. Since we assume that the motion of the binary system occurs on the equatorial plane, the third derivative of the quadrupole moment has the form $\dddot{\textbf{D}}=(\dddot{D}_x,\dddot{D}_y,\dddot{D}_z)$. Therefore,
   the quadrupole contribution to the magnetic field reduces to
   \begin{equation}
       \label{s5ae10}
       \textbf{B}_{\text{quadrupole}}=\frac{1}{6c^3L}(\dddot{D}_y\cos\iota,\dddot{D}_z\sin\iota-\dddot{D}_x\cos\iota,-\dddot{D}_y\sin\iota).
   \end{equation}
   Using, Eqs.~(\ref{s2e3}) and (\ref{s2e5}), and the relation $D_\alpha=D_{\alpha\beta}n_\beta$, it is straightforward to compute $\dddot{D}_\alpha$. Nevertheless, because the third derivative of $D_x$, $D_y$ and $D_z$ involves the first and higher derivatives of $R$ and $\psi$, it is important to remark that we use the quasi-circular approximation to simplify the expressions. Therefore, in dimensionless units, the quadrupole contribution in dimensionless units reduces to (See Appendix~\ref{A0})
   \begin{equation}
       \label{s5ae11}
       \begin{aligned}
       \textbf{B}_\text{quadrupole}&=\frac{\mu^2R^2\omega^3_s}{L}\left(\frac{\lambda_1}{m_1}+\frac{\lambda_2}{m_2}\right)\\
       &\times(\sin2\iota\cos2\psi,-\sin2\iota\sin2\psi,2\sin^2\iota\cos2\psi).
       \end{aligned}
   \end{equation} 
	For simplicity, we consider an observer located along the $x$-axis with a distance $L>>R$. Therefore, $\iota=\pi/2$ and Eqs.~(\ref{s5ae10}) and (\ref{s5ae11}) reduce to
   \begin{equation}
       \label{s5ae12}
       \begin{aligned}
       \textbf{B}_\text{dipole}&=\frac{m_1m_2(\lambda_1-\lambda_2)(1-\lambda)}{R^2L}\sin\frac{\Phi}{2}\hat{\textbf{k}},\\\\
       \textbf{B}_\text{quadrupole}&=\frac{2\mu^2R^2\omega^3_s}{L}\left(\frac{\lambda_1}{m_1}+\frac{\lambda_2}{m_2}\right)\cos\Phi \hat{\textbf{k}},
       \end{aligned}
   \end{equation}
   where we take into account the relation $\Phi=2\psi$ in the expression for $\textbf{B}_\text{quadrupole}$. Then, the total magnetic field is given by 
   \begin{equation}
       \label{s5ae13}
       \textbf{B}_\text{Total}=\textbf{B}_\text{dipole}+\textbf{B}_\text{quadrupole}.
   \end{equation}
   
   According to to Eq.~(\ref{s5ae12}), both contributions lay along the $z$-axis. Moreover, note that $\textbf{B}_\text{dipole}$ is proportional to $\sin\Phi/2$, while $\textbf{B}_\text{quadrupole}$ is proportional to $\cos\Phi$. On the other hand, it is important to point out that $\textbf{B}_\text{dipole}$ is inversely proportional to the second power of the radial separation, $R^2$, and the observer distance $L$, while the quadrupole contribution to the magnetic field $\textbf{B}_\text{quadrupole}$ is proportional not only to $R^2$ but also to the third power of the angular frequency of the source $\omega^3_s$. Equation~ (\ref{s5ae12}) also shows that the dipole contribution vanishes when $\lambda_1=\lambda_2$. Hence, as mentioned before, it is necessary to consider higher contributions to the magnetic field, i. e. the quadrupole contribution. 
   
	\begin{figure*}[t]
		\begin{center}
			\includegraphics[scale=0.35]{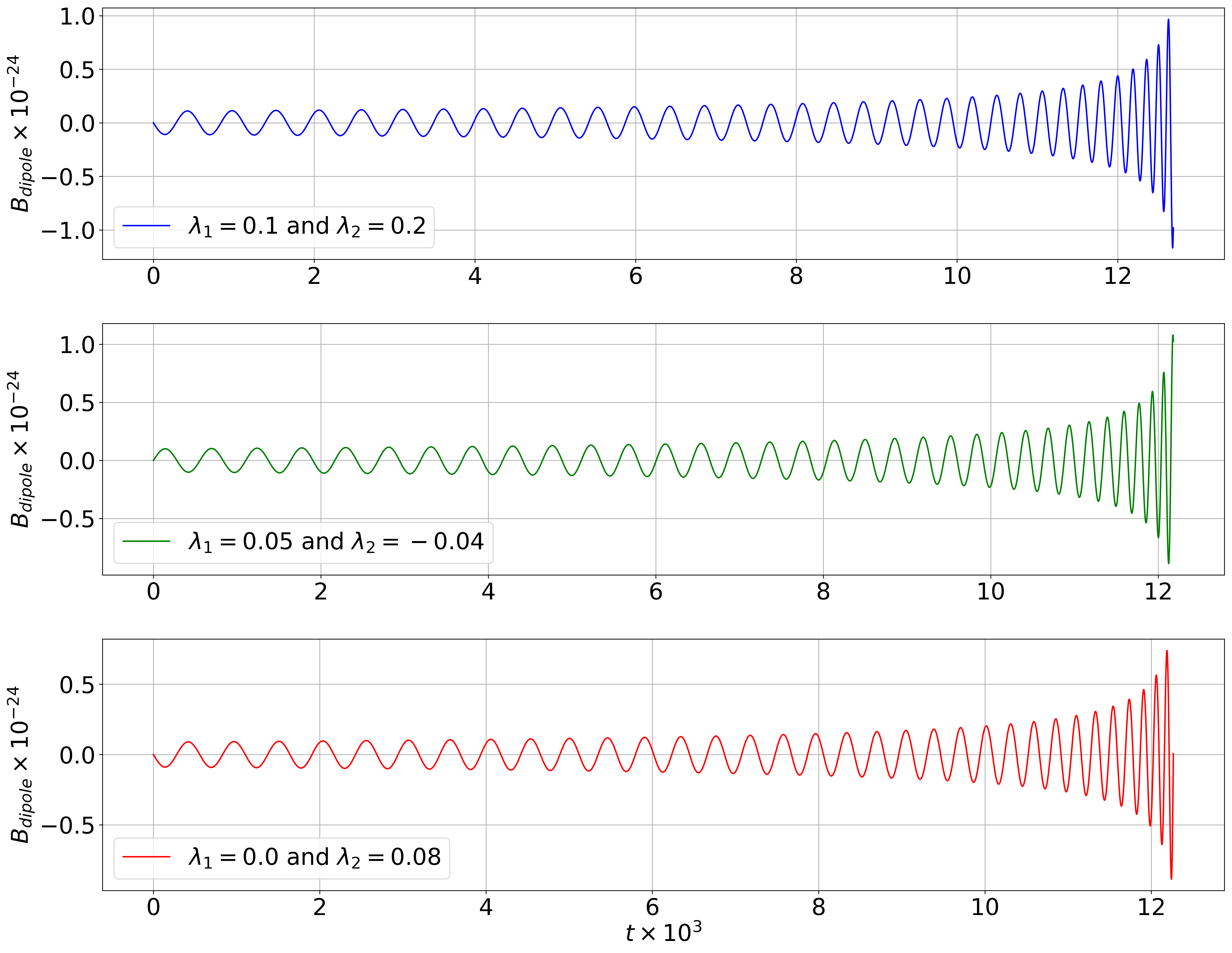}
			\caption{$\textbf{B}_\text{dipole}$ as a function of $t$ for different values of $\lambda_1$ and $\lambda_2$. For the plot we consider $m_1=m_2=1/2$, $\mu=1/4$, $R_0=20$, $\psi_0=0$ and $L=5.644\times10^{20}$, see Appendix~\ref{A0}.\label{fig5}}
		\end{center}
	\end{figure*}
   
   From the phenomenological point of view, it is possible to investigate the behavior of the EM field during the inspiral phase of a binary system formed by charged black holes by replacing the analytical expressions of Eq.~(\ref{s5e11}) in Eq.~(\ref{s5ae12}). In this sense, in Fig.~\ref{fig5}, we plot the behavior of $\textbf{B}_\text{dipole}$ as a function of time for different values of $\lambda_1$ and $\lambda_2$. We consider an observer at a distance $L=5.644\times10^{20}$ (see Appendix~\ref{A0}) along the $x$-axis. From the figure, we can see that the order of magnitude of the EM field is small, $10^{-24}$. The figure also shows how the EM field increases as the binary system approach the coalescence at the ISCO. In the case of $\lambda_1=0.1$ and $\lambda_2=0.2$, see the continuous blue line, $\textbf{B}_\text{dipole}$ oscillates between $-1\times10^{-24}$ and $1\times10^{-24}$. A similar behavior occurs when $\lambda_1=0.05$ and $\lambda_2=-0.04$. See the continuous green line. When one of the black holes does not have an electric charge, i. e. $\lambda_1=0.0$ and $\lambda_2=0.08$, the dipole contribution to the magnetic field oscillates between $-0.5\times10^{-24}$ and $0.5\times10^{-24}$. 
   
   In the figure, we can see how the presence of electric charge affects the duration time of the signal. For example, when the two black holes have $Q_1>0$ and $Q_2>0$, the inspiral phase is longer than the other two cases: two black holes with opposite charges (green line) and a binary system in which one of the black holes has none charge (red line). As mentioned above, the fact that charges with the same sign repel each other enables the binary system to interact for more time before the coalescence at the ISCO, in contrast to the case in which the black holes have an opposite electric charge, where the attraction makes the interaction shorter.  
   
	\begin{figure*}[t]
		\begin{center}
			\includegraphics[scale=0.35]{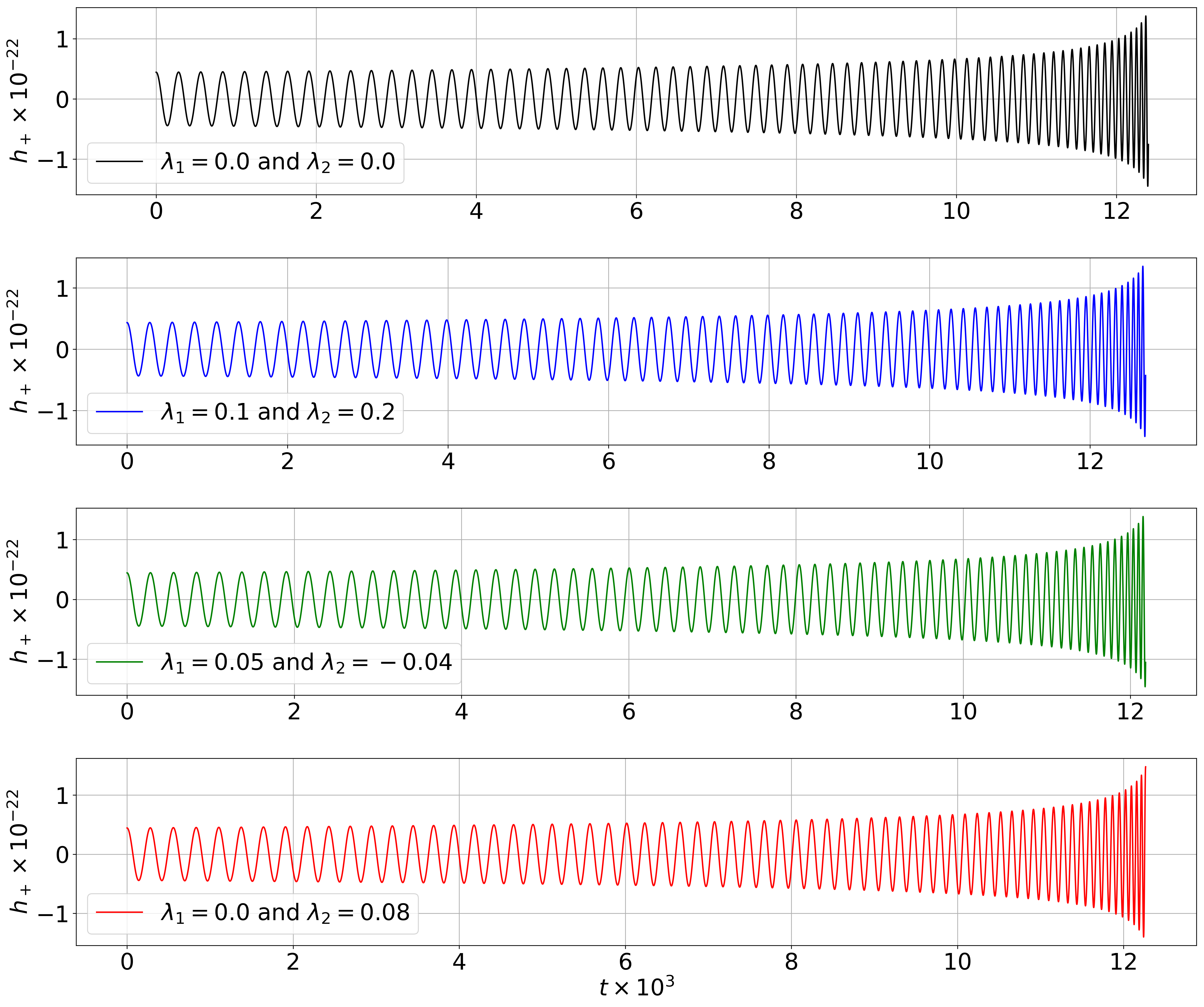}
			\caption{$h_+$ vs. $t$ for different values of $\lambda_1$ and $\lambda_2$. For the plot we consider $m_1=m_2=1/2$, $\mu=1/4$, $R_0=20$, $\psi_0=0$ and $L=5.644\times10^{20}$, see Appendix~\ref{A0}.\label{fig6}}
		\end{center}
	\end{figure*}
   
   \subsection{The gravitational wave}
   
   The plus and cross polarization of the gravitational wave of a point particle with reduced mass $\mu$ are given, in dimensionless units, by~\cite{Maggiore:2007ulw}
   \begin{equation}
        \label{s5be1}
        \begin{aligned}
        h_+&=\frac{4\mu\omega^2_s R^2}{L}\left(\frac{1+\cos^2\iota}{2}\right)\cos\Phi,\\\\
        h_\times&=\frac{4\mu\omega^2_s R^2}{L}\cos\iota \sin\Phi.
        \end{aligned}
    \end{equation}
   Where $L\rightarrow L/R_*$, $\mu\rightarrow G\mu/c^2/R_*$, $\omega_s\rightarrow\omega_s R_*/c$ and $R\rightarrow R/R_*$. See Appendix~\ref{A0}. In the case of an observer located along the $x$-axis at a distant $L$, $h_\times$ vanishes and $h_+$ takes the form
   \begin{equation}
        \label{s5be2}
        h_+=\frac{2(1-\lambda)}{\sqrt{2}L}\frac{M^{5/4}_c}{R}\cos\Phi.
   \end{equation}
   where we have considered the Keplerian law in Eq.~(\ref{A0e9}) and defined the \textit{chirp mass}, $M_c$, in dimensionless units by 
   \begin{equation}
       \label{s5be2a}
       M^5_c=\frac{1}{64}\left(\frac{4\mu}{M}\right)^{4/3}.
   \end{equation}
   
   In fig.~\ref{fig6}, we plot the GW form in the plus polarization for different values of $\lambda_1$ and $\lambda_2$ using the data from the numerical solution of Eq.~(\ref{s5e4}) and considering $L=5.644\times10^{20}$. The figure shows how $h_+$ oscillates between $-1\times10{-22}$ and $1\times10^{-22}$. Therefore, the magnitude of $h_+$ is greater than the corresponding EM wave. Furthermore, the figure also shows that the magnitude of $h_+$ increases slowly in the last part of the inspiral phase (when $10\times10^3<t<12\times10^3$) in contrast to the EM waveform, which increases its magnitude faster than $h_+$ in the same interval of time. 
   
   On the other hand, similar to the EM waveform, the electric charge affects the duration of the inspiral time before coalescence. Hence, in the case of a binary system formed by non-charged black holes, the coalescence time is longer than the cases in which $\lambda_1=0.05$ and $\lambda_2=-0.04$, or $\lambda_1=0.0$ and $\lambda_2=0.08$. Nevertheless, when $\lambda_1=0.1$ and $\lambda_2=0.2$, the coalescence time is longer than in the other cases. Once again, this behavior is due to the attraction/repulsion between electric charges.\\  
  
	\begin{figure*}[t]
	\begin{center}$
			\begin{array}{ccc}
			\includegraphics[scale=0.29]{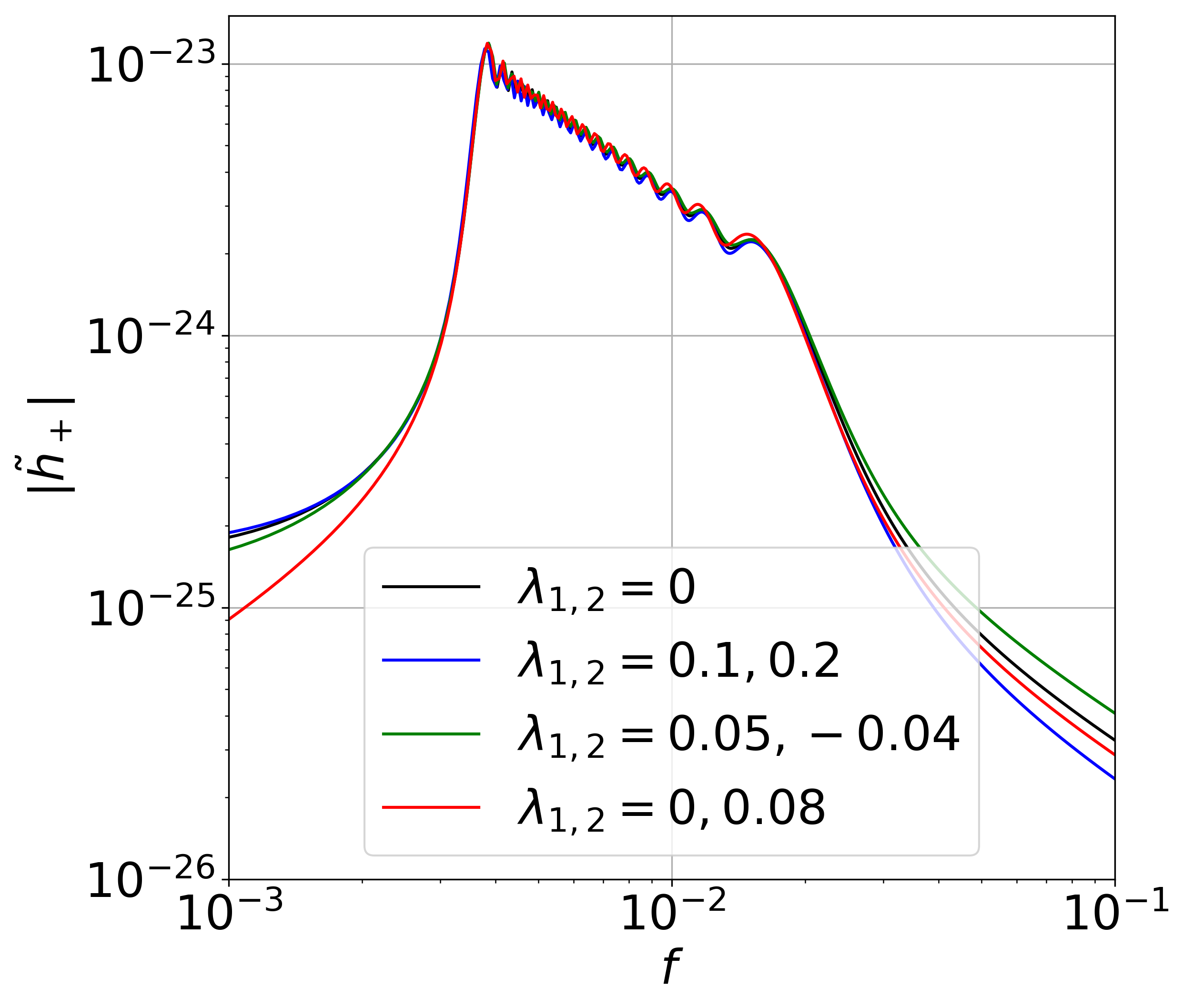} &
			\includegraphics[scale=0.29]{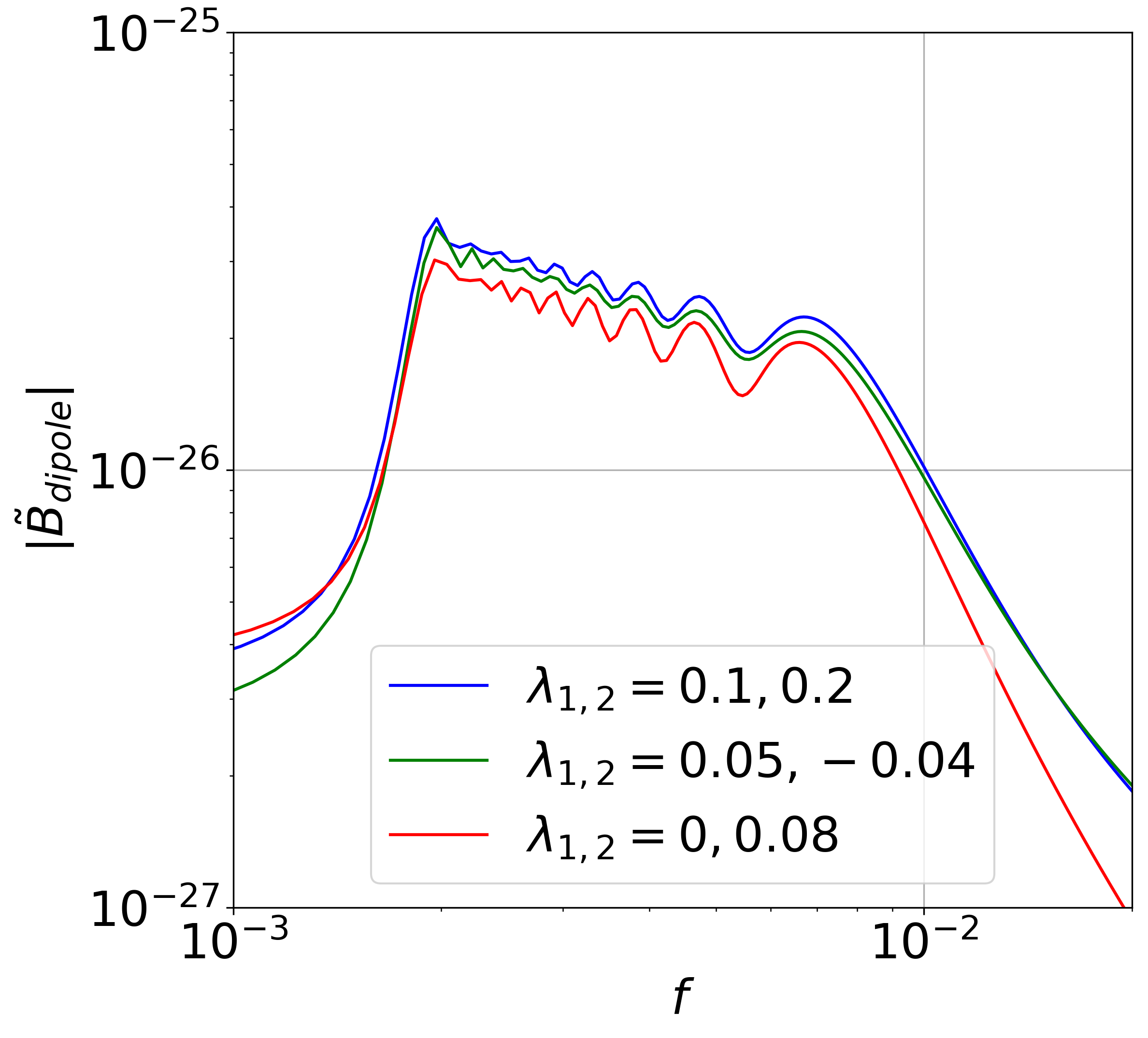}&
			\includegraphics[scale=0.29]{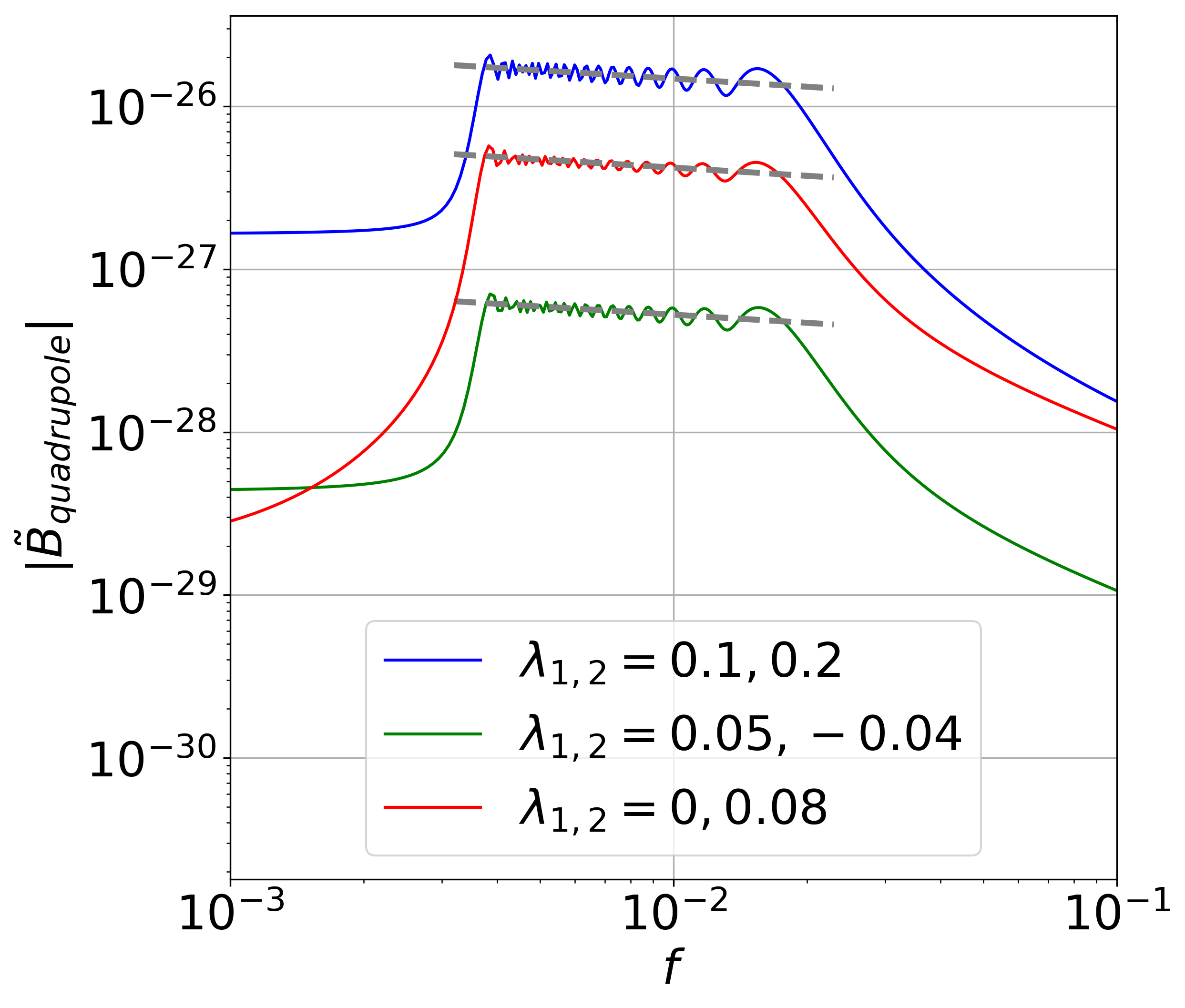}
			\\
			\includegraphics[scale=0.29]{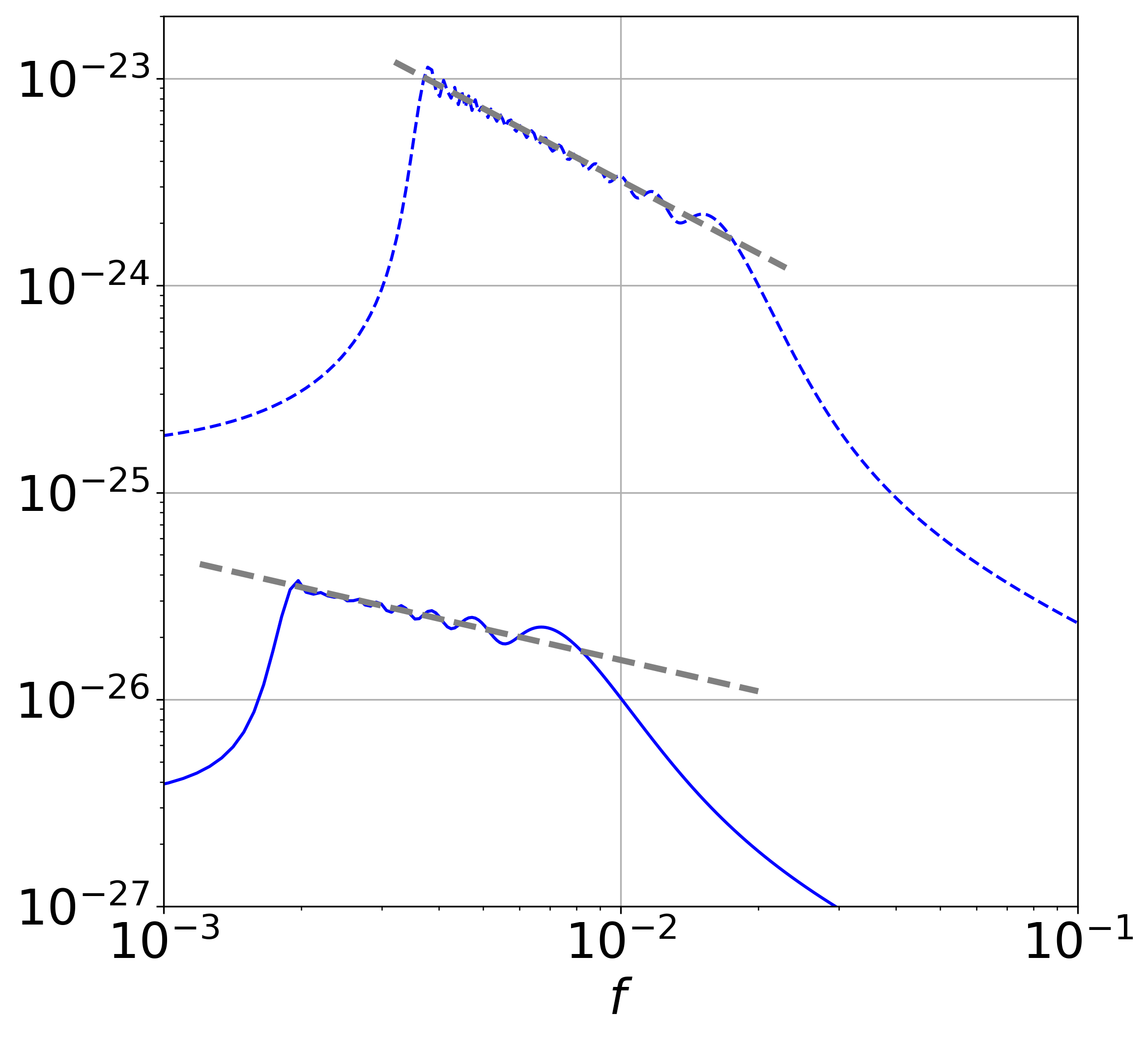}&
			\includegraphics[scale=0.29]{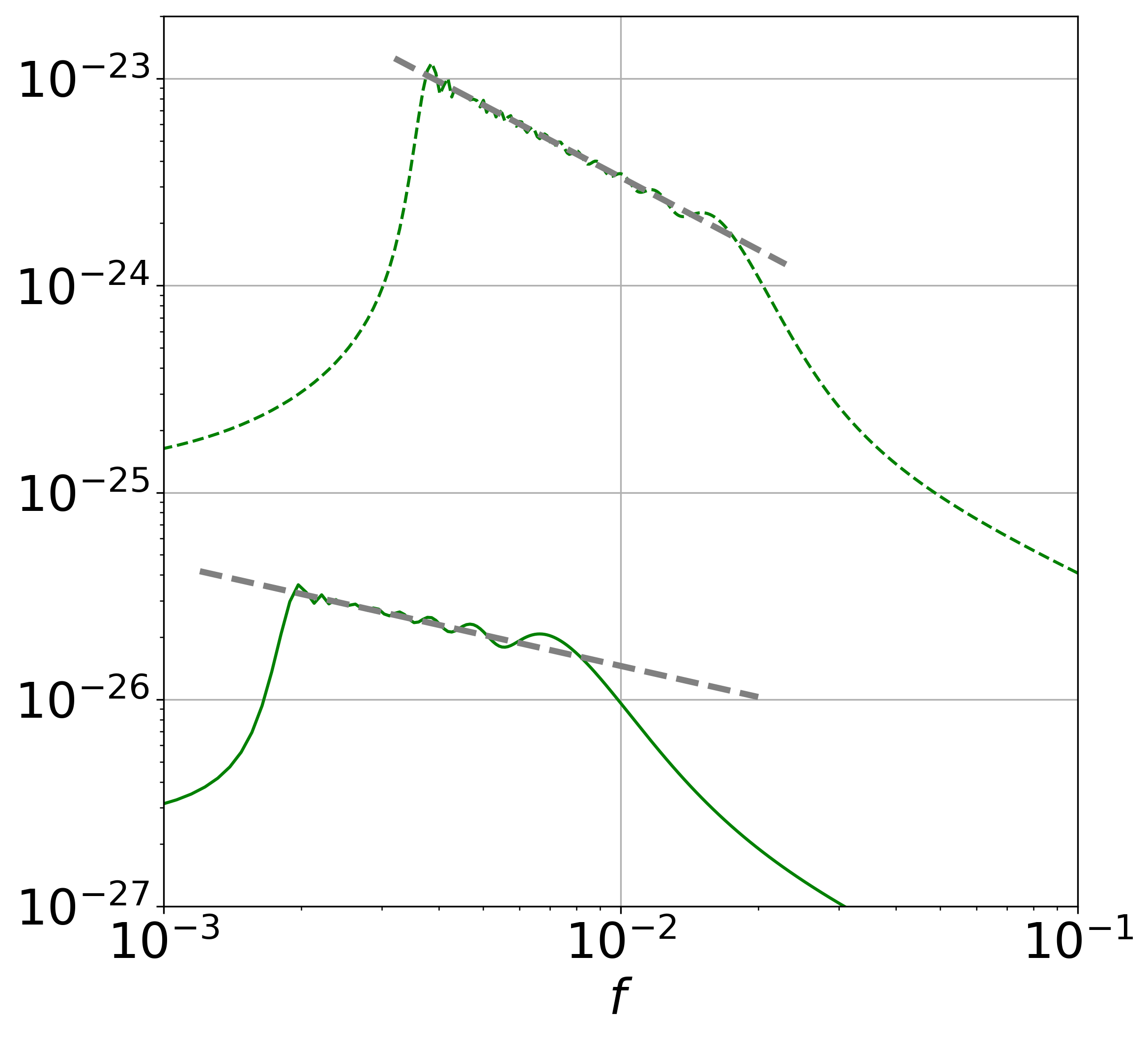}&
			\includegraphics[scale=0.29]{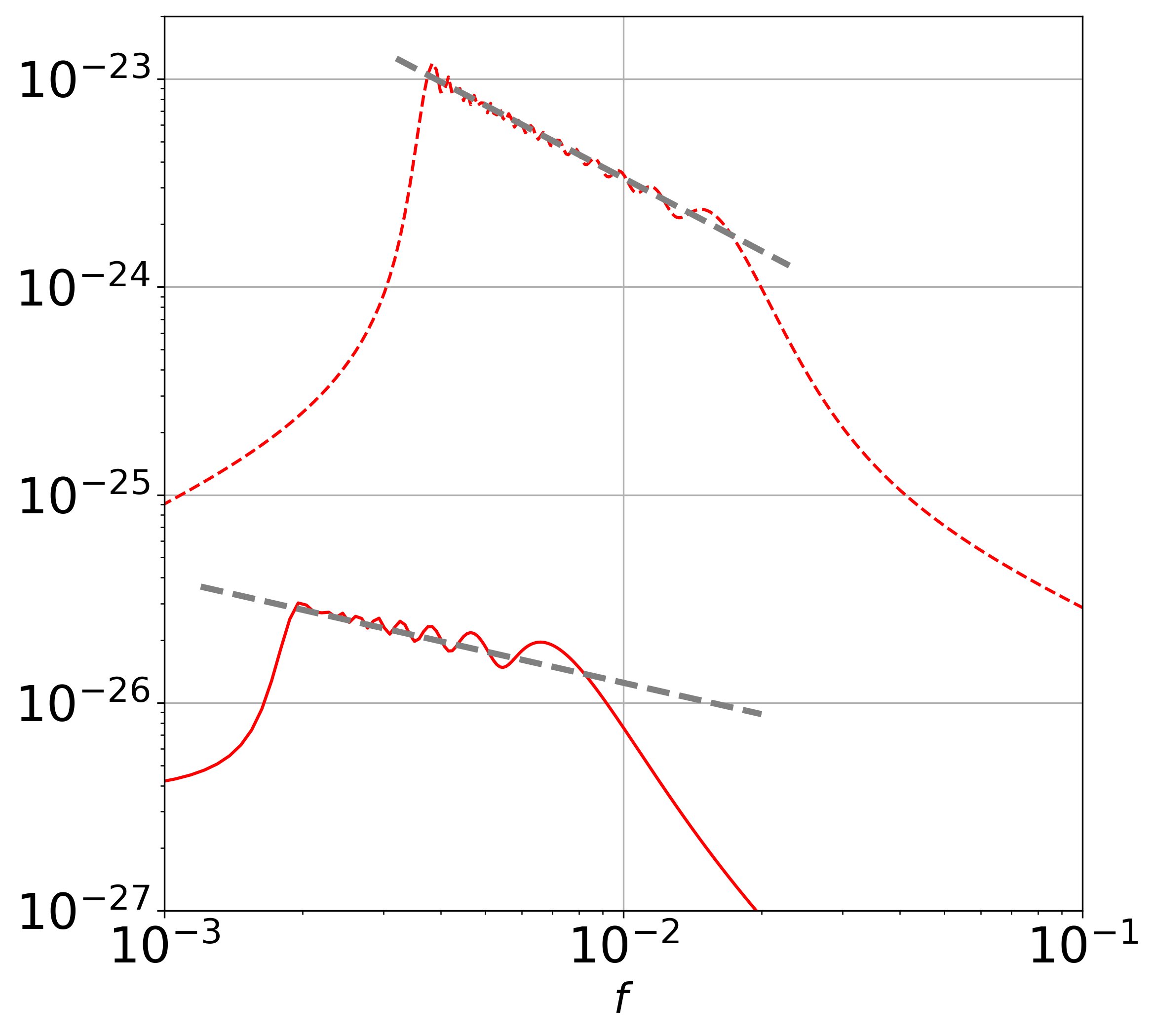}
			\end{array}$
		\end{center}
	\caption{The Fourier transform of $h_+$, $\textbf{B}_\text{dipole}$ and $\textbf{B}_\text{quadrupole}$ using the data from the numerical solution of Eq.~(\ref{s5e4}) for different values of $\lambda_1$ and $\lambda_2$. In the bottom panels, $\tilde{h}_+$ is plotted using dashed lines while $\tilde{\textbf{B}}_\text{dipole}$ with solid lines. The gray dashed lines correspond to the analytical power law. For the plot we consider $m_1=m_2=1/2$, $\mu=1/4$, $R_0=20$, $\psi_0=0$ and $L=5.644\times10^{20}$, see Appendix~\ref{A0}.\label{fig7}}
	\end{figure*}
  
   \section{Fourier transform\label{SecVII}}
   
   Under certain conditions, it is possible to express a function $g(t)$ (the ``\textit{signal}'') as a linear combination of sines and cosines with the help of the Fourier series. Furthermore, one can also represent the signal in the frequency domain, which shows the collection of frequencies constituting the function $g(t)$. The representation of $g(t)$ in the frequency domain, denoted by $\tilde{g}(f)$, is known as the Fourier transform and is a powerful tool used to analyze and obtain information regarding the evolution and behavior of the signal. In this sense, the central purpose of this section is to calculate the Fourier transform of the EM wave. To do so, we follow the ideas of Ref.~\cite{Maggiore:2007ulw}.
   
   \subsection{Gravitational wave} 
   
   In the last section, we discussed the behavior of $h_+$ as a function of time using Eq.~(\ref{s5be2}) and the data from the numerical solution of Eq.~(\ref{s5e4}). Nevertheless, to compute the Fourier transform, it is necessary to find an analytical expression. For that purpose, it is necessary to use the approximation for $R$ and $\Phi$ obtained in Sec.~\ref{SecV}. Hence, after replacing Eq.~(\ref{s5e11}) in Eq.~(\ref{s5be2}), we obtain the following expression
   \begin{equation}
   \label{s8ae1}
   h_+=A(t)\cos\Phi(t), 
   \end{equation}
   with 
   \begin{equation}
       \label{s8ae2}
       A(\tau)=\frac{(1-\lambda)^{1/2}}{2L}M^{5/4}_c\left(\frac{5}{\tau}\right)^{1/4}-\frac{(\lambda_1-\lambda_2)^2}{12L}M^{5/2}_c,
   \end{equation}
   where we have considered the approximation $\gamma R<<1$.
   It is straightforward to show that Eq.~(\ref{s8ae1}) reduces to the Eq.~(4.361) of Ref.~\cite{Maggiore:2007ulw} when $\lambda_1$ and $\lambda_2$ vanish.
   
   Following Ref.~\cite{Maggiore:2007ulw}, the Fourier transform of Eq.~(\ref{s8ae1}) is given by 
   \begin{equation}
       \label{s8ae3}
       \tilde{h}_+(f)=\int^\infty_{-\infty}dt A(t_\text{r})\cos\Phi(t_\text{r})e^{2\pi f t i}.
   \end{equation}
   Note that the integrand is evaluated at the retarded time $t_r=t-L$ (dimensionless units). Hence, after taking into account that $dt=dt_r$ and the Euler's formula, the last expression reduces to
   \begin{equation}
       \label{s8ae4}
       \tilde{h}_+(f)=\frac{1}{2}e^{2\pi f L i}\int^\infty_{-\infty} dt_r A(t_r)\left[e^{i\Phi(t_r)}+e^{-i\Phi(t_r)}\right] e^{2\pi f t_r i}.
    \end{equation}
    According to Ref.~\cite{Maggiore:2007ulw}, because  $\dot{\Phi}=\omega_\text{GW}>0$ only the term proportional to $e^{[-\Phi(t_\text{r})+2\pi f t_\text{r}]i}$ has a stationary point, while the term proportional to $e^{[\Phi(t_\text{r})+2\pi f t_\text{r}]i}$ is always oscillating fast, and integrates to a negligible small value. Hence, the Fourier transformation reduces to~\cite{Maggiore:2007ulw}
    \begin{equation}
        \label{s8ae5}
        \tilde{h}_+\simeq\frac{1}{2}e^{2\pi f L i}\int^\infty_{-\infty} dt_r A(t_r)e^{i[2\pi f t_r -\Phi(t_r)]}.
    \end{equation}
    We can compute the last integral using the stationary phase method. Hence, because $A(t_\text{r})$ varies slowly in contrast to $\dot{\Phi}$, the stationary point $t_*(f)$ can be obtained by the condition $2f\pi=\dot{\Phi}(t_*)=\omega_\text{Gw}$, which means that the largest contribution to the Fourier transform with a given $f$ is obtained for the values of $t$ for which $\omega_\text{GW}$ is equal to $2\pi f$. Therefore, after expanding the exponential in the integrand up to the second order in $(t-t_*)$, one obtains~\cite{Maggiore:2007ulw}  
    \begin{equation}
        \label{s8ae6}
        \tilde{h}_+(f)=\frac{1}{2}e^{i\Psi_+(t_*)}A(t_*)\left(\frac{2\pi}{\ddot{\Phi}(t_*)}\right)^{1/2},
    \end{equation}
    where~\cite{Maggiore:2007ulw}
    \begin{equation}
        \label{s8ae7}
        \Psi_+(t_*)=2\pi f(L+t_*)-\Phi(t_*)-\frac{\pi}{4}.
    \end{equation}
   From Eq.~(\ref{A2e19}) it is straightforward to show that 
   \begin{equation}
       \label{s8ae8}
       \begin{aligned}
       \Phi(\tau)&=\Phi_0-2(1-\lambda)^{-1/4}(5M_c)^{-5/8}\tau^{5/8}\\\\
       &+\frac{(1-\lambda)^{-3/4}(\lambda_1-\lambda_2)^2}{14\sqrt{5}}(5M_c)^{5/8}\tau^{7/8}.
       \end{aligned}
   \end{equation}
   To obtain the analytical expression for $\tilde{h}_+$, it is necessary to obtain $t_*$. To do so, we start by considering the condition $\dot{\Phi}(t_*)=\omega_\text{GW}$. Hence, from Eq.~(\ref{s8ae8}), we obtain the following relation
   \begin{equation}
       \label{s8ae10}
       \begin{aligned}
       \omega_\text{GW}&=2(1-\lambda)^{-1/4}\left(\frac{5}{256}\frac{1}{\tau_*}\right)^{3/8}M^{-5/8}_c\\\\
       &-\frac{1}{8}(1-\lambda)^{-3/4}(\lambda_1-\lambda_2)^2\left(\frac{5}{256}\frac{1}{\tau_*}\right)^{1/8}M^{5/8}_c.
       \end{aligned}
   \end{equation}
   Nevertheless, since $(\lambda_1-\lambda_2)^2<<1$, the second term Eq.~(\ref{s8ae10}) is small and we can use the approximation
   \begin{equation}
       \label{s8ae11}
       \omega_\text{GW}\approx2(1-\lambda)^{-1/4}\left(\frac{5}{256}\frac{1}{\tau_*}\right)^{3/8}M^{-5/8}_c.
   \end{equation}
   Therefore, the stationary point is given by 
   \begin{equation}
       \label{s8ae12}
       \tau_*=\frac{5}{256}(1-\lambda)^{-2/3}M^{-5/3}_c(\pi f)^{-8/3},
   \end{equation}
   which reduces to Eq.~(4.19) of Ref.~\cite{Maggiore:2007ulw} when $\lambda=0$. After replacing in Eqs.~(\ref{s8ae7}) and (\ref{s8ae8}), we obtain the following expression for the phase~\cite{Christiansen:2020pnv,Cardoso:2016olt}
   \begin{equation}
       \label{s8ae13}
       \begin{aligned}
       \Psi_+(f)&=2\pi f (L+t_\text{coal})+\frac{3}{4}(1-\lambda)^{-2/3}(M_c 8 \pi f)^{-5/3}\\
       &-\frac{5}{14}(1-\lambda)^{-4/3}(\lambda_1-\lambda_2)^2M^{3/2}_c(M_c 8 \pi f)^{-7/3}\\
       &-\frac{\pi}{4}-\Phi_0.
       \end{aligned}
   \end{equation}
   Finally, from Eq.~(\ref{s8ae11}), the second time derivative  is given by 
   \begin{equation}
       \label{s8ae14}
       \ddot{\Phi}(\tau)=\frac{192}{5}(1-\lambda)^{-1/4}\left(\frac{5}{256}\frac{1}{\tau}\right)^{11/8}M^{-5/8}_c,
   \end{equation}
   from which
   \begin{equation}
       \label{s8ae15}
       \ddot{\Phi}(\tau_*)=\frac{192}{5}(1-\lambda)^{2/3}M^{5/3}_c(\pi f)^{11/3}.
   \end{equation}
   Hence, after replacing in Eq.~(\ref{s8ae6}), the analytical expression for the Fourier transform $\tilde{h}_+$ is given by
   \begin{widetext}
   \begin{equation}
       \label{s8ae16}
       \begin{aligned}
       \tilde{h}_+(f)=\frac{1}{2L}\left[\left(\frac{5}{24}\right)^{1/2}\frac{1}{\pi^{2/3}}\right](1-\lambda)^{1/3}M^{5/6}_cf^{-7/6}e^{i\Psi_+}-\frac{1}{48L}\left[\left(\frac{5}{24}\right)^{1/2}\frac{1}{\pi^{4/3}}\right](1-\lambda)^{-1/3}(\lambda_1-\lambda_2)^2M^{5/12}_cf^{-11/6}e^{i\Psi_+}.
       \end{aligned}
   \end{equation}
   \end{widetext}
   Note that the last expression reduces to Eq.~(4.34) of Ref.~\cite{Maggiore:2007ulw} when $\lambda_1$ and $\lambda_2$ vanish.
   
   In Fig.~\ref{fig7}, we plot the Fourier transform for the electromagnetic and gravitational waves for different values of $\lambda_1$ and $\lambda_2$. In particular, the first-row left panel of the figure shows the behavior of $\tilde{h}_+$ as a function of $f$. The figure shows that $\tilde{h}_+$ behaves similarly to the Fourier transform of a binary system formed by non-charged black holes. According to Eq.~(\ref{s8ae16}), this behavior is expected by noticing that the term proportional to $f^{-7/6}$ dominates over the term proportional to $f^{-11/6}$ when $(\lambda_1-\lambda_2)^2<<1$. 
   
   \subsection{Electromagnetic wave}
   
   From Eqs.~(\ref{s5ae12}), (\ref{s5e11}), (\ref{A2e6}) and the approximation $\gamma R<<1$, the dipole contribution takes the form
   \begin{equation}
       \label{s8be1}
       B_\text{dipole}=B(\tau)\cos\left(\frac{\Phi}{2}-\frac{\pi}{2}\right),
   \end{equation}
   where
   \begin{equation}
       \label{s8be2}
       \begin{aligned}
       B(\tau)&=\frac{m_1m_2(\lambda_1-\lambda_2)}{8L}\left(\frac{5}{\tau}\right)^{1/2}\\
       &-\frac{m_1m_2(\lambda_1-\lambda_2)^3(1-\lambda)^{-1/2}}{24L}M^{5/4}_c\left(\frac{5}{\tau}\right)^{1/4}.
       \end{aligned}
   \end{equation}
   
   The Fourier transform is given by 
   \begin{equation}
       \label{s8be3}
       \tilde{B}_\text{dipole}=\int^\infty_{-\infty}dt B(t_r)\cos\left(\frac{\Phi(t_r)}{2}-\frac{\pi}{2}\right)e^{2\pi f t i}.
   \end{equation}
   Once again, note that the integrand is evaluated at the retarded time $t_r=t-L$. Hence, after taking into account that $dt=dt_r$ and the Euler's formula, the last expression reduces to\footnote{Recall that $\tau=t_\text{coal}-t$.}
   \begin{equation}
       \label{s8be4}
       \tilde{B}_\text{dipole}\simeq\frac{e^{i(2\pi fL+\pi/2)}}{2}\int^\infty_{-\infty}dt_r B(t_r)e^{i\rho(tr)}.
   \end{equation}
   Where we take into account that only the term proportional to $e^{[-\Phi(t_\text{r})+2\pi f t_\text{r}]i}$ has a stationary point and define 
   \begin{equation}
       \label{s8be5}
       \rho= 2\pi f t_r-\frac{\Phi(t_r)}{2}.
   \end{equation}
   Expanding $\rho$ up to second order around the stationary point $t_*$, we obtain the following relation
   \begin{equation}
       \label{s8be6}
       \begin{aligned}
       \rho&\approx2\pi f t_*-\frac{\Phi(t_*)}{2}+\left[2\pi f- \frac{\dot{\Phi}(t_*)}{2}\right](t_r-t_*)\\
       &-\frac{\ddot{\Phi}(t_*)}{4}(t_r-t_*)^2.
       \end{aligned}
   \end{equation}
   Hence, the stationary point for the dipole contribution can be obtained by the condition 
   \begin{equation}
       \label{s8be7}
       2\pi f=\frac{\dot{\Phi}(t_*)}{2}=\frac{\omega_\text{GW}}{2},
   \end{equation}
   from which 
   \begin{equation}
       \label{s8be8}
       \rho(t_r)\approx2\pi f t_*-\frac{\Phi(t_*)}{2}-\frac{\ddot{\Phi}(t_*)}{4}(t_r-t_*)^2,
   \end{equation}
   and
   \begin{equation}
       \label{s8be9}
       \tilde{B}_\text{dipole}=\frac{e^{i[2\pi f(L+t_*)+\pi/2-\Phi(t_*)/2]}}{\sqrt{\ddot{\Phi}(t_*)}}\int^\infty_{-\infty}dxe^{-ix^2}.
   \end{equation}
   In the last expression, we consider the change of variable 
   \begin{equation}
       \label{s8be10}
       x=\sqrt{\frac{4}{\ddot{\Phi}(t_*)}}(t_r-t_*).
   \end{equation}
   Therefore, after integration, we obtain 
   \begin{equation}
       \label{s8be11}
       \tilde{B}_\text{dipole}=e^{i\Psi_\text{dipole}}B(t_*)\left(\frac{\pi}{\ddot{\Phi}(t_*)}\right)^{1/2},
   \end{equation}
   where
   \begin{equation}
       \label{s8be12}
       \Psi_\text{dipole}=2\pi f(L+t_\text{coal})-2\pi f\tau_*+\frac{\pi}{4}-\frac{\Phi(\tau_*)}{2}.
   \end{equation}
   To obtain the last expression, we consider the relation $t_*=t_\text{coal}-\tau_*$. Now, from Eq.~(\ref{s8be7}), we obtain
   \begin{equation}
       \label{s8be13}
       \tau_*=5(1-\lambda)^{-2/3}M^{-5/3}_c(16\pi f)^{-8/3}.
   \end{equation}
   From which, after replacing in Eq.~(\ref{s8be12}), we obtain
   \begin{equation}
       \label{s8be14}
       \begin{aligned}
       \Psi_\text{dipole}&=2\pi f (L+t_\text{coal})+\frac{3}{8}(1-\lambda)^{-2/3}(M_c 16 \pi f)^{-5/3}\\
       &-\frac{5}{28}(1-\lambda)^{-4/3}(\lambda_1-\lambda_2)^2M^{3/2}_c(M_c 16 \pi f)^{-7/3}\\
       &+\frac{\pi}{4}-\frac{\Phi_0}{2}.
       \end{aligned}
   \end{equation}
   Now, from Eqs.~(\ref{s8be2}), (\ref{s8be11}), and (\ref{s8be13}), we get 
   \begin{widetext}
    \begin{equation}
        \label{s8be15}
        \tilde{B}_\text{dipole}=\frac{1}{2L}\left(\frac{5}{24}\right)^{1/2}m_1m_2(\lambda_1-\lambda_2)f^{-3/6}e^{i\Psi_\text{dipole}}-\frac{1}{24L}\left(\frac{5}{24}\frac{1}{\pi^{4/3}}\right)^{1/2}m_1m_2(\lambda_1-\lambda_2)^3(2(1-\lambda))^{-2/3}M^{5/6}_cf^{-7/6}e^{i\Psi_\text{dipole}}.
    \end{equation}
   \end{widetext}
   Finally, in the case of the quadrupole contribution, Eq.~(\ref{s5ae12}) reduces to
   \begin{equation}
       \label{s8be16}
       B_\text{quadrupole}=C(\tau)\cos\Phi(\tau),
   \end{equation}
   where we define 
   \begin{equation}
       \label{s8be17}
       \begin{aligned}
       C(\tau)&=\frac{1}{8L}(1-\lambda)^{1/4}\left(\frac{\lambda_1}{m_1}+\frac{\lambda_2}{m_2}\right)M^{25/8}_c\left(\frac{5}{\tau}\right)^{5/8}\\
       &-\frac{5}{96L}\frac{(\lambda_1-\lambda_2)^2}{(1-\lambda)^{1/4}}\left(\frac{\lambda_1}{m_1}+\frac{\lambda_2}{m_2}\right)M^{35/8}_c\left(\frac{5}{\tau}\right)^{3/8}.
       \end{aligned}
   \end{equation}
   Once again, we considered the Keplerian law in Eq.~(\ref{A0e9}) and the approximation $\gamma R<<1$.
   
   From Eqs.~(\ref{s8ae1}) and (\ref{s8be16}), it is possible to see $h_+$ and $B_\text{quadrupole}$ have the same behavior. The only difference lies in the value of the amplitudes $A(\tau)$ and $C(\tau)$. In this sense, the GW and the the quadrupole contribution of the magnetic field will have the same phase; i. e. $\Psi_\text{quadrupole}=\Psi_+$. Therefore, the Fourier transform is given by the relation
   \begin{equation}
       \label{s8be18}
        \tilde{B}_\text{quadrupole}(f)=\frac{1}{2}e^{i\Psi_+(\tau_*)}C(\tau_*)\left(\frac{2\pi}{\ddot{\Phi}(\tau_*)}\right)^{1/2},
   \end{equation}
   from which
   \begin{widetext}
   \begin{equation}
       \label{s8be19}
       \tilde{B}_\text{quadrupole}(f)=\frac{1}{2L}\left(\frac{5\pi}{6}\right)^{1/2}\left(\frac{\lambda_1}{m_1}+\frac{\lambda_2}{m_2}\right)e^{i\Psi_+(\tau_*)}\left[(1-\lambda)^{1/3}M^{10/3}_c(\pi f)^{-1/6}-\frac{(1-\lambda)^{-1/3}}{48}(\lambda_1-\lambda_2)^2M^{25/6}_c(\pi f)^{-17/6}\right].
   \end{equation}
   \end{widetext}
   
   In the first row and second panel of Fig.~\ref{fig7}, we plot the Fourier transform of the dipole contribution as a function of $f$ for different values of $\lambda_1$ and $\lambda_2$. The figure shows that $|\tilde{B}_\text{dipole}|$ is between $10^{-26}$ and $10^{-25}$, in contrast to that of the gravitational wave, where the order of magnitude goes between $10^{-24}$ and $10^{-23}$. Furthermore, similarly to $|\tilde{h}_+|$, $|\tilde{B}_\text{dipole}|$ decreases as the frequency $f$ increases. This behavior is shown clearly in the second row of Fig.~\ref{fig7}, where we plot the Fourier transform of $h_+$ and $B_\text{dipole}$ together. It is important to point out that the rate in which $|\tilde{B}_\text{dipole}|$ decreases is less than that of $|\tilde{h}_+|$, i. e. while the Fourier transform in the case of $h_+$ is dominated by a term proportional to $f^{-7/6}$, the Fourier transform of $B_\text{dipole}$ is dominated by a term proportional to $f^{-3/6}$, see Eqs.~(\ref{s8ae16}) and (\ref{s8be15}), respectively. Hence, while $|\tilde{h}_+|$ decreases with a slope of $-7/6$, $|B_\text{dipole}|$ decreases with a smaller slope of $-3/6$, see the dashed gray lines in the figure. 
   

   \section{Conclusions \label{SecVIII}}
   
   Binary systems of charged black holes have been considered by several authors~\cite{Bozzola:2019aaw, Liu:2020cds, Liu:2020vsy, Liu:2020bag, Christiansen:2020pnv, Bozzola:2020mjx, Bozzola:2021elc,Wang:2021vmi,Liu:2022cuj,Luna:2022udb}. For example, in Ref.~\cite{Bozzola:2019aaw}, the authors initiate an exploration of charged binary systems by developing an initial data formalism within the framework of the conformal transverse-traceless (Bowen-York) technique using the puncture approach and applying the theory of isolated horizons to attribute physical parameters to each black hole, such as mass, charge, and angular momentum. According to the authors, this work helps simulate the dynamical evolution of several systems. In particular, the ultrarelativistic head-on collision and the quasi-circular or eccentric inspiral and merger of two black holes~\cite{Bozzola:2020mjx, Bozzola:2021elc}. In Refs.~\cite{Liu:2020cds, Liu:2020vsy, Liu:2020bag, Liu:2022cuj}, the authors investigate the case of binaries systems with electric and magnetic charges in circular and elliptical orbits on a cone. In Ref.~\cite{Christiansen:2020pnv}, on the other hand, the authors considered the inspiral of black holes carrying not electromagnetic charge but $U(1)$ charge, which corresponds to the dark sector. They investigate how the orbital parameters evolve for dipole-dominated emission, finding that the orbit also circularises, though not efficiently, in contrast to gravitationally dominated emissions. They also investigate the modification in the gravitational waveform when the binary system contains small charges. Then, the authors combine the waveform with simplified LIGO noise and perform a matched filtering procedure, where the template bank only consists of uncharged templates. In this way, and focusing on the charges' effect on the chirp mass estimation, they found a consistent overestimation of the ``generalized'' chirp mass and a possible over- and underestimation of the actual chirp mass. 
   
   Recently, in Ref.~\cite{Luna:2022udb}, R.~Luna et al. investigated the emission of linear momentum (or ``\textit{kicks}'') produced by both gravitational and electromagnetic radiation. To do so, the authors considered the fully general-relativistic numerical evolution of quasi-circular charged black hole binaries. They also derived analytical expressions in the case of slowly moving bodies to explore, numerically, a variety of mass ratios and charge-to-mass ratios. In the case of equal masses, they found that these expressions are in excellent agreement with the observed values and that, in contrast to the vacuum case, ``\textit{kicks}'' occur in the presence of electromagnetic fields. In the case of unequal mass, the authors found that strong gravitational ``\textit{kicks}'' affect the electromagnetic ones and their magnitudes are always smaller than the gravitational ``\textit{kicks}''.  
   
   In this work, we have investigated the electromagnetic radiation generated by a binary system of charged black holes during the inspiral phase by the post-Newtonian approximation. To do so, we first compute the Keplerian motion of two point masses $m_1$ and $m_2$, with charges $Q_1$ and $Q_2$, respectively. This approximation allowed us to reduce the two-body problem to a one-body problem (with \textit{reduced mass} $\mu$) under the influence of a potential, which contains two contributions: gravitational and electric potentials, see Eq.~(\ref{s2e7}). Then, using the Lagrangian formalism, we obtained the radial separation between the two charged black holes in the system, $R$, and the angular velocity $\omega_s=\dot{\psi}$ in terms of the orbital parameters: the eccentricity, $\epsilon$, and the semi-major axis, $a$. It is significant to point out that the Kepler law changes by a factor of $(1-\lambda)^{1/2}$, which reduces to $1$ when the charges are zero. These expressions play a crucial role when obtaining analytical representations for both gravitational and electromagnetic waves.
   
   On the other hand, intending to obtain the evolution of the orbital parameters, it is necessary to discuss the gravitational and electromagnetic radiation of the system. In this sense, we follow the work of Lang Liu et al. in Ref.~\cite{Liu:2020cds}, where they consider a point-mass binary system with electric charges in a Keplerian orbit. Hence, with the help of energy conservation, Lang Lui et al. found the differential equations that describe the evolution of the semi-major axis, $a$, and the eccentricity, $\epsilon$. See Eqs.~(\ref{s4e13}) and (\ref{s4e17}), respectively. From these equations, it is possible to see that the Keplerian orbit circularizes ($\epsilon=0$) after some time, giving entrance to the inspiral phase. In this stage of the evolution of the binary system, the radial separation $R$ follows the differential equation (\ref{s5e4}). This equation contains a logarithmic function in its solution, and it is necessary to use an approximation to find an analytical expression. For this reason, we use the work of Christiansen et al., where the authors obtain analytical expressions for $R$, $\omega_s$, and $\Phi$ under the approximation $\gamma R <<1$, which means that the difference between the mass-to-charge ratio of each black hole, $\lambda_1-\lambda_2$ is much smaller than unity~\cite{Christiansen:2020pnv}. See Eq.~(\ref{s5e11}). Using this approximation, we found analytical expressions for $h_+$ and $h_\times$ configurations of the gravitational wave, the dipole and quadrupole contributions to the magnetic field, and their Fourier transforms. For simplicity, we considered an observer along the $x$-axis.
   
   Our results show that $h_+$ is formed by two terms: one proportional to $(5/\tau)^{1/4}$ and a small constant proportional to $(\lambda_1-\lambda_2)^2$. The latter is a small contribution if $(\lambda_1-\lambda_2)^2<<1$. When the binary system does not have an electric charge, $h_+$ reduces to the well-known relation, see Ref.~\cite{Maggiore:2007ulw}. However, the presence of electric charge in the binary system does affect the Keplerian orbit during the inspiral phase. For example, when the two black holes in the binary system have a positive charge, the inspiral phase will last longer than the other cases, i. e. $\lambda_1=\lambda_2=0$, $\lambda_1>0$ and $\lambda_2<0$, and $\lambda_1=0$ and $\lambda_2>0$, see Figs.~\ref{fig4} (left panel) and \ref{fig6}. As mentioned above, this behavior results from the phenomenological interaction between the charges, which repel/attract when they have the same/opposite signs. 
   
   We found a similar behavior for the EM waves. Nevertheless, while the magnitude of $h_+$ oscillates between $\pm1\times 10^{-22}$, the magnitude of the dipole contribution to the magnetic field oscillates between $\pm1\times 10^{-24}$. Therefore, the magnetic field detected by the observer will be of order $3.932\times 10^{-15}\text{Gauss}$. Moreover, it is important to remark that $|\mathbf{B}_\text{dipole}|$ is proportional to $\sin(\Phi/2)$ while $|\mathbf{B}_\text{quadrupole}|$ is proportional to $\cos\Phi$. Hence, the quadrupole contribution of the magnetic field will have the same phase as the GW. From the analytical point of view, we found that $\textbf{B}_\text{dipole}$ also has two contributions: one proportional to $(5/\tau)^{1/2}$ and the other proportional to $(5/\tau)^{1/4}$. Once again, the former will dominate because the second one, proportional to $(\lambda_1-\lambda_2)^3$, is much smaller, see Eq.~(\ref{s8be1}). The same situation occurs for $\textbf{B}_\text{quadrupole}$. There, we have two contributions proportional to $(5/\tau)^{5/8}$ and $(5/\tau)^{3/8}$, where the former also dominates. See Eq.~(\ref{s8be16}).
   
   Finally, following the ideas of Ref.~\cite{Maggiore:2007ulw}, we compute the Fourier transform for $h_+$, $\textbf{B}_\text{dipole}$ and $\textbf{B}_\text{quadrupole}$. In all the cases, we found two contributions. For example, in the GWs, we obtain two terms proportional to $f^{-7/6}$ and $f^{-11/6}$ respectively, see Eq.~(\ref{s8ae16}). However, the former term dominates if $(\lambda_1-\lambda_2)^2<<1$. Note that the exponent of the dominant term is the same as that of the binary system formed by non-charged black holes, and its behavior is similar even if we change the values of $\lambda_1$ and $\lambda_2$. Hence, the frequencies constituting the GW signal during the inspiral phase belong to the same interval in all the cases (different values of $\lambda_1$ and $\lambda_2$), and they increase as the magnitude of $\tilde{h}_+$ decreases. Our waveform model could be useful for the estimation of charge of black holes in LIGO-Virgo-KAGRA GW events.
   
   In the case of the dipole contribution to the magnetic field, we also found that its Fourier transform has two terms proportional to $f^{-3/6}$ and $f^{-7/6}$, respectively. See Eq.~(\ref{s8be15}). Once again, as in the case of $\tilde{h}_+$, the former term dominates. Furthermore, it is important to remark that $|\tilde{\textbf{B}}_\text{dipole}|$ is more sensitive to the change in the values of $\lambda_1$ and $\lambda_2$, but the frequencies constituting the EM wave always belong to the same interval. Therefore, the presence of electric charge affects the magnitude of the Fourier transform, but it does not affect the distribution of frequencies during the inspiral phase; this will be the same in all cases (different values of $\lambda_1$ and $\lambda_2$). On the other hand, note that the frequency interval in the dipole contribution of the magnetic field is different from that of the GW. This is due to the different phases, see Eqs.~(\ref{s8ae13}) and (\ref{s8be14}), and Fig.\ref{fig7}. Nevertheless, the behavior of $|\tilde{\textbf{B}}_\text{dipole}|$ as a function of $f$ is the same as in the Fourier transform of the GW, i. e. Its magnitude decreases as $f$ increases. 
   
   We also consider the quadrupole contribution. Nevertheless, as mentioned above, we include it to complete our discussion since our focus is the dipole contribution. As expected, the quadrupole contribution to the magnetic field has the same phase as the GW. That is a consequence of its proportionality of $\cos\Phi$. Furthermore, similar to the Fourier transform of the dipole and the gravitational wave, the quadruple contribution contains two terms proportional to $f^{-1/6}$ and $f^{-17/6}$, where the former is the dominant term, see Eq.~(\ref{s8be19}). The results show that the lowest contribution occurs when $\lambda_1=0.05$ and $\lambda_2=-0.04$ (when the binary system contains black holes with opposite signs) and followed by the cases where $\lambda_1=0.0$ and $\lambda_2=0.08$, and $\lambda_1=0.1$ and $\lambda_2=0.2$, respectively. See the first row, right panel of  Fig.\ref{fig7}. When compared with the $|\tilde{\textbf{B}}_\text{dipole}|$, we found that the quadrupole contribution has the same order in the latter case, i. e. when $\lambda_1=0.1$ and $\lambda_2=0.2$.

   \begin{acknowledgments}
    This work was supported by the National Key R\&D Program of China, Grant No. 2021YFC2203002. The work of C.A.B.G is supported by the President’s International Fellowship Initiative (PIFI) program of the Chinese Academy of Sciences. W. H. is supported by CAS Project for Young Scientists in Basic Research YSBR-006, NSFC (National Natural Science Foundation of China) No. 12173071, No. 12111530107 and No. 11773059, and the Strategic Priority Research Program of the CAS under Grants No. XDA15021102.
   \end{acknowledgments}

    \newpage
    \appendix
    \section{Units\label{A0}}
    In CGS units we have the following values~\cite{Misner:1973prb}
    \begin{equation}
    \label{A0e1}
    \begin{aligned}
    M_\odot&=1.989\times10^{33} \text{g},\\
    G&=6.6743\times10^{-8}\text{cm}^3\text{g}^{-1}\text{s}^{-2},\\
    c~&=2.9979\times 10^{10}\text{cm}\text{s}^{-1},\\
    k_e&=1.
    \end{aligned}
    \end{equation}
    On the other hand, the units for the charge and magnetic field are defined by 
    \begin{equation}
    \label{A0e2}
    \begin{aligned}
    &\\
    [Q]&=1\text{statcoulomb}\equiv\text{cm}^{3/2}\text{g}^{1/2}\text{s}^{-1},\\
    [B]&=1\text{Gauss}\equiv\text{cm}^{-1/2}\text{g}^{1/2}\text{s}^{-1}.
    \end{aligned}
    \end{equation}
    In Sec.~\ref{SecII} we define $\lambda=k_e\lambda_1\lambda_2/G$ in Eq.~(\ref{s2e28}). Hence, using Eqs.~(\ref{A0e1}) and (\ref{A0e2}), it is straightforward to show that $\lambda$ and $\lambda_i/\sqrt{G}=Q_i/(\sqrt{G}m_i)$ are dimensionless.
    
    In Sec.~\ref{SecV}, we discuss the quasi-circular approximation. A dimensional analysis of Eq.~(\ref{s5e1}) shows that $dR/dt$ has units of velocity. In this sense, if we multiply by the factor $1/c$, we obtain its version in geometrized units, i. e.
    \begin{equation}
    \label{A0e3}
    \begin{aligned}
    \frac{1}{c}\frac{dR}{dt}&=-\alpha\frac{R^3_*}{R^3}-\beta\frac{R^2_*}{R^2}
    \end{aligned}
    \end{equation}
    with\footnote{In the expression for $\beta$, we obtain a term of the form $\lambda_i/\sqrt{G}$. Nevertheless, since $\lambda_i$ ($i=1,2$) has the same dimension as $\sqrt{G}$, we simply write $\lambda_i$.}
    \begin{equation}
    \label{A0e4}
    \begin{aligned}
    \alpha &=\frac{16(1-\lambda)^2}{5},\\
    \beta  &=\frac{1}{3}(1-\lambda)(\lambda_1-\lambda_2)^2\left(\frac{4\mu}{M}\right)^{1/3},
    \end{aligned}
    \end{equation}
    and
    \begin{equation}
    \label{A0e5}
    R^3_*=\left(\frac{2GM}{c^2}\right)^2\left(\frac{G\mu}{c^2}\right).
    \end{equation}
    From the last expression, after a dimensional analysis, it is possible to show that $R_*$ has units of length ($\text{cm}$). In this sense, by defining the dimensionless variables $R\rightarrow\frac{R}{R_*}$ and $t\rightarrow\frac{ct}{R_*}$, Eq.~(\ref{A0e3}) reduces to 
    \begin{equation}
    \label{A0e6}
    \frac{dR}{dt}=-\frac{\alpha}{R^3}-\frac{\beta}{R^2}.
    \end{equation}
    
    Throughout the manuscript, we use dimensionless variables in plots and mathematical expressions. Therefore, it is important to explain how to convert them. First, we need to point out that we consider $m_1=m_2=m=10M_\odot$. Therefore, $R_*$ reduces to
    \begin{equation}
        \label{A0e7}
        R_*=\frac{2Gm}{c^2}=2.952\times10^6\text{cm}
    \end{equation}
    from which
    \begin{equation}
        \label{A0e8}
        \begin{aligned}
        m&\rightarrow\frac{Gm/c^2}{R_*}=\frac{1}{2}\\
        M&\rightarrow\frac{GM/c^2}{R_*}=\frac{2Gm/c^2}{R_*}=1\\
        \mu&\rightarrow\frac{G\mu/c^2}{R_*}=\frac{Gm/c^2}{2R_*}=\frac{1}{4}
        \end{aligned}
    \end{equation}
    
    In dimensionless units the Kepler law in Eq.~(\ref{s3ae4}) is given by 
    \begin{equation}
    \label{A0e9}
    \omega^2_s\rightarrow\frac{R^2_*\omega^2_s}{c^2}=\frac{(1-\lambda)}{R^3}\left(\frac{4\mu}{M}\right)^{-1/3},
    \end{equation}
    where $\omega_s=\dot{\psi}$. From Eq.~(\ref{A0e8}), note that $4\mu/M=1$.
    
    In Sec.~\ref{SecV}, we use the following relation to compute the ISCO~\cite{Wang:2021vmi} (in geometrized units)
    \begin{equation}
    \label{A0e10}
    R_{ISCO}=\frac{4\left(\frac{GM}{c^2}\right) \left(\frac{\lambda_*}{\sqrt{G}}\right)^2}{3+\frac{1}{C}+C},
    \end{equation}
    with
    \begin{equation}
    \label{A0e11}
    C=-\left[9-8\left(\frac{\lambda_*}{\sqrt{G}}\right)^2-4\sqrt{4\left(\frac{\lambda_*}{\sqrt{G}}\right)^4_*-9\left(\frac{\lambda_*}{\sqrt{G}}\right)^2+5}\right]^{1/3}.
    \end{equation}
    and 
    \begin{equation}
    \label{A0e12}
    \lambda_*=\text{min}\left[\left| \frac{m_1\lambda_1+m_2\lambda_2}{M}\right|,\left| \frac{m_2\lambda_1+m_1\lambda_2}{M}\right|\right].
    \end{equation}
    Hence, in dimensionless units, we have that 
    \begin{equation}
    \label{A0e13}
    \begin{aligned}
    R_{ISCO}&\rightarrow\frac{R_{ISCO}}{R_*}=\frac{2\left(\frac{2GM}{c^2}/R_*\right) \left(\lambda_*\right)^2}{3+\frac{1}{C}+C}\\
    &=\frac{4\lambda^2_*}{3+\frac{1}{C}+C}\left(\frac{4\mu}{M}\right)^{-1/3}.
    \end{aligned}
    \end{equation} 
    where we write $\lambda_*\rightarrow\lambda_*/\sqrt{G}$. See footnote 9. 
    
    In Sec.~\ref{SecVI}, we compute the dipole and quadrupole contributions of the EM field using Eqs.~(\ref{s5ae6}) and (\ref{s3be5}). From Fig.~\ref{fig4}, the observer is located at a distance $L$ along the direction of $\textbf{n}$, which form an angle $\iota$ with the z-axis. Therefore, $\textbf{n}=(\sin\iota,0,\cos\iota)$ and
    \begin{equation}
        \label{A0e14}
        \begin{aligned}
        \textbf{B}_\text{dipole}&=\frac{1}{c^2 L}\ddot{\textbf{p}}\times\textbf{n}=\frac{G^{3/2}m_1m_2(\lambda_1-\lambda_2/)(1-\lambda)}{c^2\sqrt{G}R^2L}\\
        &\times(-\sin\psi\cos\iota,\cos\psi\cos\iota,\sin\psi\sin\iota).
        \end{aligned}
    \end{equation}
    A dimensional analysis of the last expression shows that $[\textbf{B}_{\text{dipole}}]=\text{cm}^{-1/2}\text{g}^{1/2}\text{s}^{-1}=1\text{Gauss}$. Hence, according to Ref.~\cite{Misner:1973prb}, to express Eq.~(\ref{A0e14}) in geometrized units we divide by the following factor 
     \begin{equation}
         \label{A0e15}
         \frac{c^2}{\sqrt{G}}=3.48\times10^{24}\text{cm}/\text{Gauss}^{-1}.
     \end{equation}
     In geometrized units, the magnetic field has units of $\text{cm}^{-1}$. Hence, to express $\textbf{B}_\text{dipole}$ in dimensionless units, we multiply by $R_*$. We obtain the following expression 
     \begin{equation}
        \label{A016}
        \begin{aligned}
        \textbf{B}_\text{dipole}&\rightarrow R_*\frac{\sqrt{G}}{c^2}\textbf{B}_\text{dipole}=\frac{m_1m_2(\lambda_1-\lambda_2)(1-\lambda)}{ R^2L}\\
        &\times(-\sin\psi\cos\iota,\cos\psi\cos\iota,\sin\psi\sin\iota)
        \end{aligned}
     \end{equation}
    where $m_{1,2}\rightarrow Gm_{1,2}/(c^2R_*)$, $R\rightarrow R/R_*$, $L\rightarrow L/R_*$, $\lambda_{1,2}\rightarrow \lambda_{1,2}/\sqrt{G}$\footnote{See footnote 9.}.  
    
    The quadrupole contribution can be computed in a similar way. From the second term in Eq.~(\ref{s5ae6}) and taking into account that the binary system moves on the equatorial plane, we have that
    \begin{equation}
        \label{A017}
        \textbf{B}_\text{quadrupole}=\frac{1}{6c^3L}(\dddot{D}_y\cos\iota,-\dddot{D}_x\cos\iota,-\dddot{D}_y\sin\iota).
    \end{equation}
    Now, from Eq.~(\ref{s5ae4}) and $D_\alpha=D_{\alpha\beta}n_\beta$, we obtain 
    \begin{equation}
        \label{A018}
        \begin{aligned}
        D_x&=D_{x\beta}n_\beta=-R^2\mu^2(3\sin^2\psi-2)\left(\frac{\lambda_1}{m_1}+\frac{\lambda_2}{m_2}\right)\sin\iota,\\
        D_y&=D_{y\beta}n_\beta=\frac{3}{2}R^2\mu^2\sin2\psi\left(\frac{\lambda_1}{m_1}+\frac{\lambda_2}{m_2}\right)\sin\iota,\\
        D_z&=D_{z\beta}n_\beta=-R^2\mu^2\left(\frac{\lambda_1}{m_1}+\frac{\lambda_2}{m_2}\right)\cos{\iota}.
        \end{aligned}
    \end{equation}
    From which
    \begin{equation}
        \label{A019}
        \begin{aligned}
        \dddot{D}_x&\approx12\mu^2R^2\omega^3_s\sin2\psi\left(\frac{\lambda_1}{m_1}+\frac{\lambda_2}{m_2}\right)\sin\iota,\\
        \dddot{D}_y&\approx-12\mu^2R^2\omega^3_s\cos2\psi\left(\frac{\lambda_1}{m_1}+\frac{\lambda_2}{m_2}\right)\sin\iota\\
        \dddot{D}_z&\approx0
        \end{aligned}
    \end{equation}
   At this point, it is important to point out that the expressions for $\dddot{D}_x$, $\dddot{D}_y$ and $\dddot{D}_z$ involve the first and higher time derivatives of $R$ and $\psi$. In this sense, and following Ref.\cite{Maggiore:2007ulw}, one can use the quasi-circular approximation to neglect these terms and simplify the derivatives as Eq.~(\ref{A019}). Thus, the quadrupole contribution reduces to
   \begin{equation}
       \label{A020}
       \begin{aligned}
       \textbf{B}_\text{quadrupole}&=\frac{\mu^2R^2\omega^3_s}{c^3L}\left(\frac{\lambda_1}{m_1}+\frac{\lambda_2}{m_2}\right)\\
       &\times(-\sin2\iota\cos2\psi,-\sin2\iota\sin2\psi,2\sin^2\iota\cos2\psi).
       \end{aligned}
   \end{equation}
   Once again, a dimensional analysis of the last expression shows that
   \begin{equation}
       \label{A021}
       [\textbf{B}_\text{quadrupole}]=\text{cm}^{-1/2}\text{g}^{1/2}\text{s}^{-1}=\text{Gauss}.
   \end{equation}
   Therefore, in dimensionless units, we obtain
    \begin{equation}
       \label{A022}
       \begin{aligned}
       \textbf{B}_\text{quadrupole}&\rightarrow R_*\frac{\sqrt{G}}{c^2}\textbf{B}_\text{quadrupole}=\frac{\mu^2R^2\omega^3_s}{L}\left(\frac{\lambda_1}{m_1}+\frac{\lambda_2}{m_2}\right)\\
       &\times(\sin2\iota\cos2\psi,-\sin2\iota\sin2\psi,2\sin^2\iota\cos2\psi).
       \end{aligned}
   \end{equation}
   where $m_{1,2}\rightarrow Gm_{1,2}/(c^2R_*)$, $\mu\rightarrow G\mu/(c^2R_*)$, $R\rightarrow R/R_*$, $L\rightarrow L/R_*$, $\lambda_{1,2}\rightarrow \lambda_{1,2}/\sqrt{G}$ and $\omega_s\rightarrow R_*\omega_s/c$. 
   
   In fig.~\ref{fig4}, we plot the behavior of $\textbf{B}_\text{dipole}$ as a function of $t$. For the figures, we consider an observer located at a distance $L=540\text{Mpc}$. In \text{CGS} units we have that
   \begin{equation}
        \label{A023}
        \begin{aligned}
        540\text{Mpc}\times&\frac{1\times10^6\text{pc}}{1\text{Mpc}}\times\frac{3.0857\times10^{16}\text{m}}{1\text{pc}}\times\frac{100\text{cm}}{1\text{m}}\\
        &=1.6663\times10^{27}\text{cm}
        \end{aligned}
    \end{equation}
    Hence, in dimensionless units, $L$ reduces to 
    \begin{equation}
        \label{A024}
        \begin{aligned}
        L\rightarrow \frac{L}{R_*}&= \frac{1.6663\times 10^{27}\text{cm}}{636037.2339\text{cm}}= 5.6441\times 10^{20}.
        \end{aligned}
    \end{equation}
    
    In Sec.~\ref{SecVI}, we discuss the GW radiated by the binary system. There are two polarization for the gravitational radiation: plus and cross, which are given by the following relations~\cite{Maggiore:2007ulw},
    \begin{equation}
        \label{A025}
        \begin{aligned}
        h_+&=\frac{1}{L}\frac{4G\mu \omega^2_s R^2}{c^4}\left(\frac{1+\cos^2\iota}{2}\right)\cos\Phi,\\
        h_\times&=\frac{1}{L}\frac{4G\mu \omega^2_s R^2}{c^4}\cos\iota \sin\Phi.
    \end{aligned}
    \end{equation}
    A dimensional analysis shows that $h_+$ and $h_\times$ are dimensionless. Nevertheless, to obtain the same expression in dimensionless variables it is necessary to rewrite the common factor in the following way
    \begin{equation}
        \label{A026}
        \frac{1}{L}\frac{4G\mu \omega^2_s R^2}{c^4}=\frac{1}{L/R_*}\left(\frac{G\mu}{c^2R_*}\right)\left(\frac{R_*\omega^2_s}{c}\right)^2\left(\frac{R}{R_*}\right)^2.
    \end{equation}
    Hence, Eq.~(\ref{A025}), in dimensionless units, reduces to 
    \begin{equation}
        \label{A027}
        \begin{aligned}
        h_+&=\frac{\mu\omega^2_s R^2}{L}\left(\frac{1+\cos^2\iota}{2}\right)\cos\Phi\\
        h_\times&=\frac{\mu\omega^2_s R^2}{L}\cos\iota \sin\Phi.
        \end{aligned}
    \end{equation}
	\section{$\epsilon$ as a function of $E$ and $L$\label{A1}}
	From the conservation of angular momentum and energy, Eqs.~(\ref{s2e15}) and (\ref{s2e24}), respectively, we have
	\begin{equation}
	\label{A1e1}
	\begin{aligned}
	\dot{R}&=\sqrt{\frac{2E}{\mu}-\frac{L^2}{\mu^2 R^2}-\frac{2\mathcal{U}}{\mu}},\\
	\dot{\psi}&=\frac{L}{\mu R^2}.
	\end{aligned}
	\end{equation}
    The last two equations represent the derivative of $R$ and $\psi$ respect to $t$. Nevertheless, it is possible to express $\psi$ as a function of $R$. To do so, we recall that
    \begin{equation}
    \label{A1e2}
    \dot{\psi}=\frac{d\psi}{dt}=\frac{d\psi}{dR}\dot{R}.
    \end{equation}
    Hence, using again the conservation of energy and angular momentum, the last expression reduces to 
    \begin{equation}
    \label{A1e3}
    \frac{d\psi}{dR}=\frac{L}{\mu R^2\dot{R}}=\frac{L}{\mu R^2\sqrt{\frac{2E}{\mu}-\frac{L^2}{\mu^2 R^2}-\frac{2\mathcal{U}}{\mu}}}.
    \end{equation}
    After integration, the last equation takes the form 
    \begin{equation}
    \label{A1e4}
    \psi=\psi_0+\int^R_{R_0}\frac{dR'}{{R'}^2\sqrt{\frac{2\mu E}{L^2}-\frac{2\mu\mathcal{U}}{L^2}-\frac{1}{{R'}^2}}}.
    \end{equation}
    Then, changing the variable to $u=1/R'$, with $du=-(1/{R'}^2)dR'$, the integral reduces to
    \begin{equation}
    \label{A1e5}
    \psi=\psi_0-\int^u_{u_0}\frac{du}{\sqrt{\frac{2\mu E}{L^2}+\frac{2\mu\kappa u}{L^2}-u^2}},
    \end{equation}
    where we consider $\mathcal{U}=-\kappa u$. The integral in Eq.~(\ref{A1e5}) is of the form~\cite{Goldstein:1980}
    \begin{equation}
    \label{A1e6}
    \int\frac{du}{\sqrt{a+b u + w u^2}}=\frac{1}{\sqrt{-w}}\arccos\left(-\frac{b+2w u}{\sqrt{b^2-4aw}}\right).
    \end{equation}
    After comparison with Eq.~(\ref{A1e5}), we have that
    \begin{equation}
    \label{A1e7}
    \begin{array}{ccc}
    a=\frac{2\mu E}{L^2},&b=\frac{2\mu\kappa}{L^2},&w=-1.
    \end{array}
    \end{equation}
    Thus, we obtain 
    \begin{equation}
    \label{A1e8}
    b^2-4aw=\left(1+\frac{2EL^2}{\mu\kappa^2}\right)\left(\frac{2\mu\kappa}{L^2}\right)^2.
    \end{equation}
    Substituting into Eq.~(\ref{A1e6}) and after integration, we get
    \begin{equation}
    \label{A1e9}
    \psi=\psi_0-\arccos\frac{\frac{L^2 u}{\mu\kappa}-1}{\sqrt{1+\frac{2EL^2}{\mu\kappa^2}}},
    \end{equation}
    from which
    \begin{equation}
    \label{A1e10}
    \frac{1}{R}=\frac{\mu\kappa}{L^2}\left[1-\sqrt{1+\frac{2EL^2}{\mu\kappa^2}}\cos{(\psi-\psi_0)}\right].
    \end{equation}
    Hence, according to Eq.~(\ref{s2e40}), the eccentricity is defined as 
    \begin{equation}
    \label{A1e11}
    \epsilon=\sqrt{1+\frac{2EL^2}{\mu\kappa^2}}.
    \end{equation}

    \section{Quasi-circular approximation ($\gamma R<<1$)\label{A2}}
    To obtain $R$ as a function of $\tau$, we start by defining $u=\tau/\tau_0$. Hence, after using Eq.~(\ref{s5e10}), we obtain the relation 
    \begin{equation}
    \label{A2e1}
    u^{1/4}=\frac{R}{R_0}\left(1-\frac{4\gamma R}{5}\right)^{1/4}
    \left(1-\frac{4\gamma R_0}{5}\right)^{-1/4}.
    \end{equation}
    Then, after solving for $R/R_0$ and considering the case $\gamma R<<1$, the last expression takes the form
    \begin{equation}
    \label{A2e2}
    \frac{R}{R_0}=u^{1/4}\left(
    1+\frac{\gamma R}{5}-\frac{\gamma R_0}{5}\right).
    \end{equation}
    where, we neglect second order terms such as $\gamma^2 R_0R/25$. In the last expression, the term $\gamma R/5$ can be expressed in terms of $u$. Hence, from Eq.~(\ref{A2e1}) and the approximation $\gamma R<<1$, it is straightforward to show that
    \begin{equation}
    \label{A2e3}
    \frac{\gamma R}{5}\approx\frac{\gamma R_0}{5}u^{1/4}.
    \end{equation}
    Finally, Eq.~(\ref{A2e2}), takes the form~\cite{Christiansen:2020pnv}
    \begin{equation}
    \label{A2e4}
    \frac{R}{R_0}=u^{1/4}\left[1-\frac{\gamma R_0}{5}\left(1-u^{1/4}\right)\right],
    \end{equation}
    which reduces to Eq.~(4.25) of Ref.~\cite{Maggiore:2007ulw} when $\gamma=0$. 
    
    On the other hand, the expression for the initial radial separation $R_0$ can be obtained from Eq.~(\ref{s5e10}). Thus, after setting $R=R_0$ and using the approximation $\gamma R<<1$, we obtain 
    \begin{equation}
    \label{A2e5}
    R_0=(4\alpha\tau_0)^{1/4}\left[1+\frac{\gamma R_0}{5}\right].
    \end{equation}
    The term $\gamma R_0/5$ can be expressed in terms of $\tau_0$ using Eq.~(\ref{s5e10}). Hence, with  $\gamma R<<1$, it is easy to show that 
    \begin{equation}
    \label{A2e6}
    \frac{\gamma R_0}{5}\approx\frac{\gamma}{5}(4\alpha\tau_0)^{1/4},
    \end{equation}
    from which, Eq.~(\ref{A2e5}) reduces to~\cite{Christiansen:2020pnv} 
    \begin{equation}
    \label{Ase7}
    R_0=(4\alpha\tau_0)^{1/4}\left[1+\frac{\gamma(4\alpha\tau_0)^{1/4}}{5}\right].
    \end{equation}
    
    Now, to compute the equation for $\omega_s$, we use the Kepler law. Hence, from Eq.~(\ref{A0e9}), we obtain   
    \begin{equation}
    \label{A2e8}
    \gamma R = \delta \omega^{-2/3}_s, 
	\end{equation}
	where 
	\begin{equation}
	\label{A2e9}
	\delta=\frac{5}{48}\frac{(\lambda_1-\lambda_2)^2}{(1-\lambda)^{2/3}}\left(\frac{4\mu}{M}\right)^{2/9}.    
	\end{equation}
	Now, from Eqs.~(\ref{A2e1}) and (\ref{A2e8}), we solve for $\omega_s/\omega_0$ to obtain
	\begin{equation}
	\label{A2e10}
	\frac{\omega_s}{\omega_0}=u^{-3/8}\left(1-\frac{3\delta\omega^{-2/3}_s}{10}+\frac{3\delta\omega^{-2/3}_0}{10}\right),
	\end{equation}
	where we use the approximation $\tilde{\gamma}\tilde{R}<<1$ and neglect second order terms. Note that the term with $\omega_s$ can be expressed in terms of $\omega_0$ and $u$ using Eq.~(\ref{A2e3}), i. e.
	\begin{equation}
	\label{A2e11}
	\frac{3\delta\omega^{-2/3}_s}{10}\approx\frac{3\delta\omega^{-2/3}_0}{10}u^{1/4},
	\end{equation}
	from which Eq.~(\ref{A2e10}) takes the form ~\cite{Christiansen:2020pnv}
	\begin{equation}
	\label{A2e12}
	\frac{\omega_s}{\omega_0}=u^{-3/8}\left[1+\frac{3}{10}\delta\omega^{-2/3}_0(1-u^{1/4})\right].
	\end{equation}
	The expression for $\omega_0$ can be obtained from Eqs.~(\ref{s5e10}) and (\ref{A2e6}), we obtain
    \begin{equation}
    \label{A2e13}
    \omega_0=\left(\frac{3\sigma}{8\tau_0}\right)^{3/8}\left(1-\frac{3}{10}\delta\omega^{-2/3}_0\right),
    \end{equation}
    with 
    \begin{equation}
    \label{A2e16}
    \sigma=\frac{5}{24}(1-\lambda)^{-2/3}\left(\frac{4\mu}{M}\right)^{-4/9}.
    \end{equation}
    Nevertheless, from Eq.~(\ref{A2e6}), we know that 
    \begin{equation}
    \label{A2e14}
    \frac{3}{10}\delta\omega^{-2/3}_0\approx \frac{3}{10}\delta\left(\frac{8\tau_0}{3\sigma}\right)^{1/4}.
    \end{equation}
    Therefore, Eq.~(\ref{A2e13}) reduces to~\cite{Christiansen:2020pnv}
    \begin{equation}
    \label{A2e15}
    \omega_0=\left(\frac{3\sigma}{8\tau_0}\right)^{3/8}\left[1-\frac{3}{10}\delta\left(\frac{8\tau_0}{3\sigma}\right)^{1/4}\right].
    \end{equation}
    
    Finally, from the relation~\cite{Maggiore:2007ulw},
    \begin{equation}
    \label{A2e17}
    \Phi=2\int^{t}_{t_0}\omega_s dt'=-\int^{\tau}_{\tau_0}\omega_{GW}d\tau',
    \end{equation}
    where $\psi=\Phi/2$ and $\omega_{GW}=2\omega_s$. Hence, after using Eqs.~(\ref{A2e12}) and (\ref{A2e14}), one obtains~\cite{Christiansen:2020pnv}
    \begin{equation}
    \label{A2e18}
    \begin{aligned}
    \Phi&=\frac{16}{5}\tau_0\left(\frac{8\tau_0}{3\sigma}\right)^{-3/8}\left[1-u^{5/8}-\frac{3\delta}{14}\left(\frac{8\tau_0}{3\sigma}\right)^{1/4}(1-u^{7/8})\right],
    \end{aligned}
    \end{equation}
    which can be expressed as
    \begin{equation}
        \label{A2e19}
        \Phi=\Phi_0-\frac{16}{5}\tau_0\left(\frac{8\tau_0}{3\sigma}\right)^{-3/8}\left[u^{5/8}-\frac{3\delta}{14}\left(\frac{8\tau_0}{3\sigma}\right)^{1/4}u^{7/8}\right],
    \end{equation}
    where 
    \begin{equation}
        \label{A2e20}
        \Phi_0=\frac{16}{5}\tau_0\left(\frac{8\tau_0}{3\sigma}\right)^{-3/8}\left[1-\frac{3\delta}{14}\left(\frac{8\tau_0}{3\sigma}\right)^{5/8} \right]
    \end{equation}
    is the value of $\Phi$ at the moment of coalescence $\tau_0$. Equation (\ref{A2e19}) reduces to Eq.~4.30 of Ref.~\cite{Maggiore:2007ulw} when $\lambda_1$ and $\lambda_2$ vanish.

	\end{document}